\documentclass[11pt]{article}
\pdfoutput=1

\usepackage{amssymb,amsthm}
\usepackage{mathtools}
\usepackage[totalwidth=17cm,totalheight=24cm]{geometry}
\usepackage{graphicx}
\usepackage{feynmp}
\newcommand{\R}{{\mathbb R}}

\newcommand{\im}{{\rm i }}

\newcommand{\la}{\langle}
\newcommand{\ra}{\rangle}

\newtheorem{lemma}{Lemma}

\newtheorem{definition}{Definition}

\newcommand\be{\begin{eqnarray}}
\newcommand\ee{\end{eqnarray}}

\begin{document}

 \unitlength = 1mm
 
 \begin{fmffile}{assoc}

\title{Colour-Kinematics duality and the Drinfeld double \\ of the Lie algebra of diffeomorphisms}
\author{Chih-Hao Fu and Kirill Krasnov \\ {}\\
{\small \it School of Mathematical Sciences, University of Nottingham, NG7 2RD, UK} } 
\date{v2: December 2016}
\maketitle
\begin{abstract} Colour-kinematics duality suggests that Yang-Mills (YM) theory possesses some hidden Lie algebraic structure. So far this structure has resisted understanding, apart from some progress in the self-dual sector. We show that there is indeed a Lie algebra behind the YM Feynman rules. The Lie algebra we uncover is the Drinfeld double of the Lie algebra of vector fields. More specifically, we show that the kinematic numerators following from the YM Feynman rules satisfy a version of the Jacobi identity, in that the Jacobiator of the bracket defined by the YM cubic vertex is cancelled by the contribution of the YM quartic vertex. We then show that this Jacobi-like identity is in fact the Jacobi identity of the Drinfeld double. All our considerations are off-shell. Our construction explains why numerators computed using the Feynman rules satisfy the colour-kinematics at four but not at higher numbers of points. It also suggests a way of modifying the Feynman rules so that the duality can continue to hold for an arbitrary number of gluons. Our construction stops short of producing explicit higher point numerators because of an absence of a certain property at four points. We comment on possible ways of correcting this, but leave the next word in the story to future work. 
\end{abstract}

\section{Introduction}

The Color-Kinematics duality \cite{Bern:2008qj} is a statement that the tree-level on-shell scattering amplitudes of Yang-Mills (YM) theory can be written as a sum over cubic graphs, with the contribution of each graph being a product of the group structure constants (colour), kinematic numerator depending on the helicities and momenta of the particles being scattered (kinematics), as well as the product of propagators, see below for a review. Moreover, the statement is that the kinematic numerators satisfy the same Jacobi-type identity as the products of the group structure constants. The assumption that YM amplitudes admit an expression satisfying the color-kinematics duality implies new relations between amplitudes \cite{Bern:2008qj}, now commonly referred to as the BCJ relations. Further, \cite{Bern:2008qj} conjectured that the graviton scattering amplitudes can be obtained from the YM ones by a simple squaring procedure, once a set of YM numerators satisfying the color-kinematics duality has been identified. 

To some extent, the colour-kinematics duality can be explained by embedding YM into string theory, as a low energy limit of the open string. One then finds that the BCJ relations follow from the monodromy properties of the open string scattering amplitude \cite{BjerrumBohr:2009rd}, \cite{Stieberger:2009hq}, \cite{Mafra:2009bz}, \cite{Tye:2010dd}, \cite{BjerrumBohr:2010hn}. Moreover, a similar embedding of gravity as a low energy limit of the closed string proves the squaring procedure that leads from YM to graviton amplitudes \cite{Tye:2010dd}, \cite{BjerrumBohr:2010hn}. Thus, the squaring procedure that allows one to obtain graviton amplitudes from the colour-kinematic dual form of the YM amplitudes is to a large extent equivalent to the famous KLT relation \cite{Kawai:1985xq} between the amplitudes of the two theories. An explicit form of the YM kinematic numerators can be obtained via the pure spinor formalism \cite{Mafra:2011kj}. 

At the same time, a Lagrangian origin of the colour-kinematic dual form of the YM amplitudes remains to a large extent obscure. Some understanding comes by considering the MHV amplitude sector \cite{Monteiro:2011pc}, see also \cite{BjerrumBohr:2012mg}. It then becomes clear that for such amplitudes, the Feynman rules of Yang-Mills theory do lead directly to amplitudes in a colour-kinematic dual form. The Jacobi identity satisfied by the kinematic numerators receives the interpretation of that of the Lie algebra of area preserving diffeomorphisms of a certain 2-dimensional space \cite{Monteiro:2011pc}. While this result clearly points in the direction of some Lie algebra being also behind the general YM amplitude, little is understood in this direction beyond the MHV case. 

The purpose of this paper is to report some observations that strengthen the expectation that there is some infinite-dimensional Lie algebra behind the colour-kinematics duality of YM theory. We do not claim to have completely understood a Lagrangian origin of this duality, but facts we present clearly point in the direction of the Lie algebra of diffeomorphisms, i.e. the Lie algebra of vector fields with their Lie bracket, as being behind the duality. Our considerations extend those of  \cite{Monteiro:2011pc} beyond the MHV amplitudes. Moreover, our approach also simplifies the MHV story. 

It is known that at 4 points the YM Feynman rules directly lead to amplitudes satisfying the colour-kinematics duality; considerations in \cite{Bern:2008qj} start from this fact. We start by observing that there is also an off-shell understanding of the 4-point case. Thus, we shall see that the cubic vertex of YM theory defines a certain anti-symmetric bracket on vector fields, and that this bracket does not satisfy the Jacobi identity. We then observe that raison d'\^etre of the quartic YM vertex is to correct for this failure of the Jacobi identity to be satisfied. Thus, once the quartic vertex is taken into account, a certain version of the Jacobi identity can be established. All our considerations are completely off-shell. 

Equipped with knowledge of a certain off-shell version of the Jacobi identity as follows from the YM Feynman rules, we search for a Lie-algebraic explanation, and find it in the so-called Drinfeld double of the Lie algebra of diffeomorphisms. Thus, we observe that a certain twist of this Drinfeld double (which is by construction a Lie algebra) is behind the Jacobi-like identity that we derive from the YM Feynman rules. 

We also put to use the Drinfeld double interpretation at higher points. The Jacobi identity at 4 points implies a partial cancelation in the sum of numerators at higher points. But the cancelation is only incomplete and, as is well-known, numerators following from the Feynman rules do not satisfy the colour-kinematics duality at higher points. However, we sketch how this could be corrected by adding new higher point interactions. Still, we can claim only a partial success in our quest for understanding the colour-kinematics. This is because in order for our  mechanism to work the YM quartic vertex must be representable as a product of two cubic vertices, which is not the case. Thus, there is still some key element missing from our story. We finish with some speculations as to what this can be, but leave the next step to future work. 

The organisation of this paper is as follows. We start by reminding the reader the statement of the colour-kinematics duality, as well as the related statement that gravity is the square of YM. Then in Section \ref{sec:bracket} we write down the YM Feynman rules. We also introduce what we call the YM bracket in this Section. This is a bracket on vector fields that follows from the YM Feynman rules, and which does not satisfy the Jacobi identity. However, this bracket is closely related to the Lie bracket on vector fields, as we show. In this Section we also explain why the colour-kinematics duality works at the level of Feynman rules for the self-dual sector of the theory. The basic reason is that in this sector the YM bracket coincides with the Jacobi satisfying Lie bracket. We compute the Jacobiator of the YM bracket in Section \ref{sec:jacobiator}. The main result of this section is that the YM bracket satisfies a version of the Jacobi identity, once the 4-valent YM vertex is taken into account. We review some basic facts about Drinfeld doubles in Section \ref{sec:drinfeld}. Most of the material here is standard, except for our description of the twist of the double by a symmetric tensor. This is the main ingredient that we need for the application to YM theory. We describe the Drinfeld double of the algebra of diffeomorphisms in Section \ref{sec:DD}. The main result in this section is the interpretation of the Jacobi-like identity discovered from the YM Feynman rules at 4 points as the Jacobi identity of the twisted Drinfeld double. We then apply our methods to five point amplitudes. We see that the Jacobi identity established at four points implies a partial cancelation in the sum of kinematic numerators. But this is incomplete cancelation, and the numerators following from the YM Feynman rules do not satisfy colour-kinematics. However, our Lie algebraic interpretation suggests a simple way to correct for this. We see that certain parts of the Lie algebra Jacobi identity at higher points are missing from what follows from the Feynman rules. To get colour-kinematic dual numerators one simply has to add the missing parts. However, there is a stumbling block in doing this which is already seen at 4 points. The last section is a discussion of how this difficulty may be overcome. 

\section{Colour-Kinematics duality}

There are many good presentations of the colour-kinematics duality, to which we refer the reader for more details. We will only need some basic information about this form of writing the YM scattering amplitudes. We will mainly follow the original paper \cite{Bern:2008qj}, with differences in notation. 

\subsection{Scattering amplitudes}

The YM scattering amplitude 
\be\label{amp}
A^{a_1,\ldots, a_n}(k_1,\epsilon_1;\ldots;k_n,\epsilon_n)
\ee
is an object that depends on (null $k_i^2\equiv(k_i k_i) =0$) momenta $k_i, i=1,\ldots,n$ of $n$ gluons being scattered, as well as on the polarisation vectors $\epsilon_i$. The polarisation vectors are null $\epsilon_i^2=0$ and transverse $(\epsilon_i k_i)=0$. The amplitude has free Lie algebra indices $a_i$, allowing one to distinguish the colour of the gluons that are scattered. As usual, using crossing symmetry it is assumed that all particles are incoming, so that the momentum conservation reads $\sum_i k_i=0$. 

The tree-level scattering amplitude is obtained by summing over all tree-level Feynman graphs with $n$ external legs, with the propagators for the external legs amputated. The polarisations $\epsilon_i$ are inserted into the external legs. Each vertex contributes according to the Feynman rules, see below, while each internal line contributes a factor of $\im/k^2$, where $k$ is the momentum on that line, as well as a factor of Kronecker delta for the Lie algebra indices. 

\subsection{The statement of colour-kinematics duality}

The statement of duality consists of several sub statements. First, it says that it is possible to write (\ref{amp}) as a sum of cubic graphs only. This part of the statement is to some extent trivial, because one can always write the contributions from the quartic vertices in terms of some effective cubic graphs, see below on how to do this in practice. Thus, a part of the colour-kinematics duality statement is that one can write (\ref{amp}) as
\be\label{sum-graphs}
A^{a_1,\ldots, a_n}(k_1,\epsilon_1;\ldots;k_n,\epsilon_n) = \sum_{t\in \,cubic\,\,\, trees} {\rm (kinematic\,\,\, factor)_t} \prod_v f^{a_v b_v c_v} \prod_e \delta^{a_e b_e} \frac{1}{k_e^2}.
\ee
Here $v$ is the set of cubic vertices of the graph, and $e$ is the set of internal edges. The momentum $k_e$ is that on the internal edge $e$, as follows from the momentum conservation. It is a sum of a certain subset of the external momenta $k_i$. The quantity $f^{abc}$ is the Lie algebra structure constant. The Lie algebra indices are labelled as follows. First, we have 3 Lie algebra indices $a_v, b_v, c_v$ at each cubic vertex $v$. Second, we have a pair of indices $a_e, b_e$ at each internal line $e$. The internal line indices are to be thought of as being associated with the ends of the line. For example, all internal lines can be oriented in some arbitrary way, and then each Lie algebra index on the internal line is either the source of the line index or the target index. The Lie algebra indices of the structure constant are to be taken as follows. They correspond to lines emanating from the vertex. If a line is an internal line, then the Lie algebra index of the corresponding end of the line should be used. If a line is an external line, then the corresponding index from the set $a_1,\ldots,a_n$ should be used. All Lie algebra indices associated to internal lines should be summed over in (\ref{sum-graphs}). Finally, the kinematic factor in (\ref{sum-graphs}) depends on the graph topology, as well as on all the external data $k_i, \epsilon_i$. It is important to emphasise that the kinematic factor does not have to be local in any sense. Thus, it is a quantity that depends on all the external momenta and all the polarisations. Finally, our convention is that we have set the YM coupling constant to unity (it is easy to reconstruct it by counting the vertices). We have also omitted some factors of the imaginary unit, which are again reconstructible by counting.

As we have already said, a representation (\ref{sum-graphs}) can always be achieved starting from the Feynman rules, but there are many ways of doing this. The non-trivial part of the colour-kinematics duality statement is that it is possible to choose the kinematic factors in such a way as to satisfy an analog of the Jacobi identity. Thus, it is clear that in (\ref{sum-graphs}) the kinematic factors as well as the propagators can be stripped off, and then each graph carries the following colour factor
\be
{\rm (colour)}_t = \prod_v f^{a_v b_v c_v} \prod_e \delta^{a_e b_e},
\ee
where as before the sum over colour indices associated with all internal edges is assumed. These colour factors then clearly satisfy the following Jacobi identity
\bigskip
\be   \label{Jacobi}
 0=\quad  \parbox{20mm}{\begin{fmfgraph*}(20,20) 
  \fmftop{i1,i2}
  \fmfbottom{o1,o2}
  \fmflabel{1}{i1}
   \fmflabel{2}{i2}
    \fmflabel{4}{o1}
     \fmflabel{3}{o2}
\fmf{plain}{i1,v1}
\fmf{plain}{i2,v1}
\fmf{plain}{v1,v2}
  \fmf{plain}{v2,o1}
  \fmf{plain}{v2,o2}
   \end{fmfgraph*}}\,\,   \quad +   \quad \parbox{20mm}{\begin{fmfgraph*}(20,20) 
  \fmftop{i1,i2}
  \fmfbottom{o1,o2}
  \fmflabel{2}{i1}
   \fmflabel{3}{i2}
    \fmflabel{4}{o1}
     \fmflabel{1}{o2}
\fmf{plain}{i1,v1}
\fmf{plain}{i2,v1}
\fmf{plain}{v1,v2}
  \fmf{plain}{v2,o1}
  \fmf{plain}{v2,o2}
   \end{fmfgraph*}}\,\, \quad+ \quad \,\, \parbox{20mm}{\begin{fmfgraph*}(20,20) 
  \fmftop{i1,i2}
  \fmfbottom{o1,o2}
  \fmflabel{3}{i1}
   \fmflabel{1}{i2}
    \fmflabel{4}{o1}
     \fmflabel{2}{o2}
\fmf{plain}{i1,v1}
\fmf{plain}{i2,v1}
\fmf{plain}{v1,v2}
  \fmf{plain}{v2,o1}
  \fmf{plain}{v2,o2}
   \end{fmfgraph*}}
\ee
\bigskip

\noindent The 3 graphs here can also be a part of a larger graph. The statement of the colur-kinematics duality is then that it is possible to choose the kinematic factors ${\rm (kinematic\,\,\, factor)_t}$ so as to satisfy the same identity (\ref{Jacobi}) as is satisfied by the colour factors. 

\subsection{Gravity as square of YM}

Part of the importance of the colour-kinematic dual presentation (\ref{sum-graphs}) of the YM amplitudes is that, once such a presentation is known, it is easy to obtain graviton scattering amplitudes. Thus, gravitons being spin two particles, their polarisation vectors can be represented as squares of spin one polarisation vectors $h_{\mu\nu}(k) = \epsilon_\mu(k)\epsilon_\nu(k)$, where $\epsilon_\mu(k)$ is a YM polarisation vector as is appropriate for a gluon of momentum $k$. We remind the reader that for a massless particle in 4 dimensions there are two possible polarisations both for YM and for gravitons. Thus, for either polarisation, the graviton polarisation vectors is representable as the square of that of YM theory. 

With the understanding that graviton polarisations are squares of those in YM theory, the statement of gravity being a square of YM theory is as follows. The graviton scattering amplitudes can be written in a form similar to (\ref{sum-graphs}), with the colour factors for each graph being replaced by another copy of the kinematic factor
\be\label{square}
{\cal M}(k_1,h_1;\ldots;k_n,h_n) = \sum_{t\in \,cubic\,\,\, trees} {\rm (kinematic\,\,\, factor)_t} \times {\rm (kinematic\,\,\, factor)_t} \prod_e \frac{1}{k_e^2}.
\ee
Once again, this formula misses the factor of Planck mass $M_p$ to the power of the number of vertices, as well as some factors of the imaginary unit, which are easy to reconstruct. 

\subsection{The puzzle of colour-kinematics}

As we have already mentioned in the introduction, the fact that gravity is a square of YM theory, at least at the level of on-shell scattering amplitudes, is to some extent explained by embedding both theories into string theory. Gravity is then the closed string, while YM is an open string, and two copies of the open string (disk) give the closed string (sphere).

From this perspective, the duality statement (\ref{sum-graphs}) with the kinematic factors satisfying the Jacobi identity (\ref{Jacobi}) is more puzzling than the statement that gravity equals YM squared. Indeed, the colour-kinematics duality says that Yang-Mills theory itself is already a square, with its amplitudes being formed from one copy of the kinematics and one copy of the colour factors. This statement can no longer be explained by the open/closed string duality, as it says that in a certain sense the open string is already is a square. The colour-kinematics duality thus seems more intriguing than its application to gravity (\ref{square}), and this is why in this paper we seek understanding of (\ref{sum-graphs}) rather than (\ref{square}). 

\section{Feynman rules, YM bracket and the self-dual sector}
\label{sec:bracket}

The main point of this section is to establish notations. The novel aspects are as follows. We point out that there is some freedom in choosing the gauge-fixing term. This freedom is parametrised by an anti-symmetric tensor. Availability of this freedom strengthens our later interpretation of structures arising in YM theory in terms of the Drinfeld double. 

Another novel aspect is our way of writing the cubic vertex (\ref{assoc}), which makes prominent the role of the Lie algebra of vector fields with its Lie bracket. We also introduce what we refer to as the YM bracket in this section. This is an anti-symmetric operation on vector fields that comes out from the YM cubic vertex. This YM bracket does not satisfy the Jacobi identity (and this is why the statement of colour-kinematics is non-trivial). In the next section we will compute its Jacobiator. 

Finally, we provide a simple explanation for why the colour-kinematic duality follows from Feynman rules for MHV amplitudes. This provides a simplified version of the story in \cite{Monteiro:2011pc}.

\subsection{The Lagrangian}

The YM Lagrangian is
\be
{\cal L}_{\rm YM}= \frac{1}{4} (F^a_{\mu\nu})^2.
\ee
Expanded around the trivial background (zero gauge field) this gives
\be\label{L3-4}
{\cal L}^{(2)} = \frac{1}{2} (\partial_\mu A_\nu^a)^2 - \frac{1}{2} \partial_\nu A_\mu^a \partial^\mu A^{\nu a}, \\ \nonumber
{\cal L}^{(3)} = f^{abc} A_\mu^b A_\nu^c \partial^\mu A^{\nu a}, \\ \nonumber
{\cal L}^{(4)} = \frac{1}{4} f^{abc} A_\mu^b A_\nu^c f^{aef} A^{\mu e} A^{\nu f}.
\ee
The quadratic part of the Lagrangian, when gauge-fixed so as to cancel the second term, gives rise to the usual Feynman gauge propagator. 

\subsection{Gauge-fixing freedom}

As we have just mentioned, the usual gauge-fixing is to add a multiple of $(\partial^\mu A_\mu^a)^2$ to the Lagrangian so as to cancel the unwanted term in ${\cal L}^{(2)}$. However, there is some freedom in the gauge-fixing, which we now exploit. Let $C^{\mu\nu}$ be an anti-symmetric tensor, which we assume to be independent of space-time coordinates. We can then consider the following quantity
\be\label{gauge-fix}
H^a_{\rm C} = \partial^\mu A_\mu^a -\frac{1}{2} f^{abc} C^{\mu\nu} A^b_\mu A^c_\nu,
\ee
where the pre factor in front of the second term is for future convenience. Let the gauge-fixing procedure add to the Lagrangian $(1/2)(H^a_{\rm C})^2$. This gives rise to the standard gauge-fixed kinetic term (and thus the standard Feynman gauge propagator $\im/k^2$). However, this also contributes to the qubic and quartic vertices
\be\label{L3-4-gf}
{\cal L}^{(3)}_{\rm g.f.} = -\frac{1}{2} f^{abc} C^{\mu\nu} A_\mu^b A_\nu^c \partial^\rho A_{\rho}^a, \\ \nonumber
{\cal L}^{(4)}_{\rm g.f.} =  \frac{1}{8} f^{abc} C^{\mu\nu} A_\mu^b A_\nu^c f^{aef} C^{\rho\sigma} A_\rho^e A_\sigma^f.
\ee
Below we shall see that the gauge-fixing freedom (\ref{gauge-fix}) makes the developed below interpretation in terms of Drinfeld double more compelling. 

\subsection{The cubic vertex}

The cubic part of the Lagrangian (\ref{L3-4}) gives rise to 6 terms in the cubic vertex. These are most conveniently written in the momentum space, and using placeholders. Thus, instead of writing $V^{abc}_{\mu_1 \mu_2 \mu_3}(k_1,k_2,k_3)$ for the vertex factor, we write the contraction $V^{abc}_{\mu_1 \mu_2\mu_3}(k_1,k_2,k_3) A_1^{\mu_1 a} A_2^{\mu_2 b} A_3^{\mu_3 c}$. We assume that all the momenta are incoming, so that the derivative gives a factor of $\im k_\mu$. There is an additional factor of $\im$ coming from in front of the action. Further, it is convenient to strip the colour, and introduce placeholders with only a space-time index. Thus, we will write our placeholders as $A_\mu^a = t^a \xi_\mu$. Then 
\be
V^{abc}_{\mu_1 \mu_2 \mu_3}(k_1,k_2,k_3) A_1^{\mu_1 a} A_2^{\mu_2 b} A_3^{\mu_3 b} = f^{abc} t_1^a t_2^b t_3^c \,v_3(\xi_1,\xi_2,\xi_3),
\ee
where the kinematic factors are
\be\label{v3}
v_3(\xi_1,\xi_2,\xi_3) = (\xi_2 k_3)(\xi_1 \xi_3)- (\xi_1 k_3)(\xi_2 \xi_3) \\ \nonumber +(\xi_3 k_1)(\xi_2 \xi_1)- (\xi_2 k_1)(\xi_3 \xi_1)+(\xi_1 k_2)(\xi_3 \xi_2)- (\xi_3 k_2)(\xi_1\xi_2) .
\ee
Here $(\xi\eta):=\xi^\mu \eta_\mu \equiv \eta^{\mu\nu} \xi_\mu \eta_\nu$ is the Minkowski metric product of one-forms $\xi_\mu, \eta_\mu$. The quantity $v_3(\xi_1,\xi_2,\xi_3)$ has the following properties. It is cyclically symmetric $v_3(\xi_1,\xi_2,\xi_3)=v_3(\xi_2,\xi_3,\xi_1)$. It is also anti-symmetric with respect to exchange of any two arguments $v_3(\xi_1,\xi_2,\xi_3)=-v_3(\xi_2,\xi_1,\xi_3)$. It should be kept in mind that each field $\xi$ comes with its associated momentum $k$, so when vector fields are exchanged, so must be the momenta. Below we shall interpret the object (\ref{v3}) as a cochain in Lie algebra cohomology. 

\subsection{Lie bracket}

Because we have the Minkowski metric in our disposal, we can identify the space of 1-forms on the manifold with the space of vector fields. We will use this identification everywhere, thinking about objects $\xi_\mu$ as either vector fields or forms, as is convenient in each particular context. We will only differentiate between 1-forms and vector fields where this is important. 

Let us consider the Lie bracket of two vector fields
\be
[\xi_1,\xi_2] := (\xi_1 \partial) \xi_2 - (\xi_2 \partial)\xi_1.
\ee
Importantly, it satisfies the Jacobi identity
\be
[\xi_1,[\xi_2,\xi_3]] + [\xi_2,[\xi_3,\xi_1]]+[\xi_3,[\xi_1,\xi_2]]=0.
\ee
For our purposes, it will be more convenient to write everything in momentum space. Replacing the partial derivative with the corresponding momentum vector (and omitting a factor of imaginary unit) we have
\be\label{Lie}
[\xi_1,\xi_2] = (\xi_1 k_2) \xi_2-(\xi_2 k_1) \xi_1.
\ee
This is the expression for the Lie bracket that will be used on most occasions. 

Using the momentum conservation, it is not hard to see that the kinematic factor $v_3(\xi_1,\xi_2,\xi_3)$ can be written in terms of the Lie bracket in the following way
\be\label{assoc}
v_3(\xi_1,\xi_2,\xi_3) = ([\xi_1,\xi_2] \xi_3) +  ([\xi_2,\xi_3] \xi_1)+ ([\xi_3,\xi_1] \xi_2).
\ee
We find it striking that the cubic vertex of YM can be written using the flat metric and the Lie bracket of vector fields. This suggests that the Lie algebra of vector fields, which is the Lie algebra of the group of diffeomorphisms on our manifold, has something to do with the structure of the gauge theory. 

We will work out the contribution from the quadratic vertex, as well as those from the $C$-dependent gauge-fixing terms below, after we introduce some additional concepts from Lie algebra cohomology theory. 

\subsection{The new bracket}

As we have seen above, the YM cubic vertex defines a certain kinematic factor $v_3(\xi_1,\xi_2,\xi_3)$, which can be viewed as completely anti-symmetric tensor evaluated on 3 vector fields. We would like to interpret this kinematic factor as a new bracket. Thus, we give it a sense of direction, and view it as the result of a certain new bracket of vector fields $\xi_1,\xi_2$, later contracted with $\xi_3$
\be\label{cochain}
[,]_{YM}: \Lambda^2 TM \to TM,  \qquad ([\xi_1,\xi_2]_{YM} \xi_3):=v_3(\xi_1,\xi_2,\xi_3).
\ee

To write an explicit expression for the new bracket, we can use  the momentum conservation $k_3=-k_1-k_2$ so that only the momenta $k_1,k_2$ appear. We get
\be\label{bracket}
[\xi_1,\xi_2]_{YM} = 2(\xi_1 k_2) \xi_2- 2(\xi_2 k_1) \xi_1 + (k_1-k_2)(\xi_1\xi_2) + (\xi_1 k_1)\xi_2-(\xi_2 k_2)\xi_1.
\ee
This operation is anti-symmetric, and thus maps the anti-symmetric second power of the tangent bundle to the tangent bundle, as indicated. However, unlike the Lie bracket, the bracket just introduced does not satisfy Jacobi identity, as we will explicitly compute below. It is not obvious from the expression (\ref{bracket}) that the left-hand-side of (\ref{cochain}) is totally anti-symmetric, but this is made obvious by the expression (\ref{assoc}). 

An important property of the new bracket, obvious from its expression via (\ref{assoc}) is its symmetry with respect to the inner product on the Lie algebra given by the metric
\be\label{symmetry}
([\xi_1,\xi_2]_{YM} \xi_3)=([\xi_2,\xi_3]_{YM} \xi_1)=([\xi_3,\xi_1]_{YM} \xi_2).
\ee
This is a property useful in explicit computations, such as the one of the Jacobiator below.

\subsection{Colour-Kinematics duality for the YM self-dual sector}

Here we will give a simple explanation for why the colour-kinematics duality is true for the self-dual sector of YM theory. The basic observation here is that in the self-dual sector, the YM bracket as defined above coincides with the Lie bracket, which does satisfy the Jacobi identity. This establishes the colour-kinematics duality in the self-dual sector, as was also done in \cite{Monteiro:2011pc} in a related, but apparently more involved fashion. 

We now outline the main idea of our argument. It easy to show using the spinor methods, that for polarisation vectors of gluons of the same helicity, the following statements are true. First, as for any polarisation vectors, these polarisations are transverse $(\epsilon k)=0$. Second, the product of any two such polarisation vectors is zero $(\epsilon_1 \epsilon_2)=0$. This would not be true if we dealt with polarisation vectors corresponding to two different helicities. Finally, and most importantly, the Lie bracket of two such polarisation vectors gives a vector with similar properties, i.e. it is transverse and orthogonal to any other polarisation vector of the same helicity
\be\label{SD-prop}
([\epsilon_1,\epsilon_2] (k_1+k_2))=0, \qquad ([\epsilon_1,\epsilon_2] \epsilon_3)=0.
\ee
The first of these properties follows from general considerations, while the second needs a simple computation to be proved. The first property follows from the fact that the Lie bracket preserves the space of transverse vector fields, i.e. the Lie bracket of two transverse vector fields is again transverse. This is a well-known fact, which can also be confirmed by a simple computation. 

We now sketch a proof of the second of the properties in (\ref{SD-prop}). Let us take the helicity in question to be negative. In our conventions this means that the corresponding helicity spinor is given by
\be\label{spin-helicity}
\epsilon_{AA'} = \frac{q_A k_{A'}}{(1q)},
\ee
where $(1q)\equiv 1^A q_A$ is the spinor contraction, and $q_A$ is the negative helicity reference spinor. The quantity $k_{A'}$ is the momentum spinor such that the gluon null momentum is written as $k_{AA'}=k_A k_{A'}$. The Lie bracket of two such polarisation vectors is then computed as
\be\label{spinor-bracket}
[\epsilon_1,\epsilon_2]_{AA'} = \frac{q_B 1_{B'} 2^B 2^{B'}}{(1q)} \frac{q_A 2_{A'}}{(2q)} - \frac{q_B 2_{B'} 1^B 1^{B'}}{(2q)} \frac{q_A 1_{A'}}{(1q)} = q_A \frac{[12]}{(1q)(2q)} \left( 1_E 1_{A'} + 2_E 2_{A'} \right) q^E.
\ee
Here $[21]\equiv 2^{A'} 1_{A'}$ is the primed spinor contraction. We use the notation $k_{1 A} \equiv 1_A$, and similarly for the primed spinors. The most important thing about the bracket (\ref{spinor-bracket}) to be noticed is that the result is again proportional to the reference spinor $q_A$. This immediately implies that the bracket is orthogonal to any other helicity vector, which is the second property in (\ref{SD-prop}). Let us also compute the projection of the bracket (\ref{spinor-bracket}) onto the momentum vector $(1+2)_{AA'} $, to check the first property (for which we gave an independent argument above). We have
\be
[\epsilon_1,\epsilon_2]_{AA'} (1+2)^{AA'} = \frac{[12]}{(1q)(2q)} (1+2)_{EA'} q^E (1+2)^{AA'} q_A = 0.
\ee
The last equality follows because we have the product of a primed spinor $(1+2)_{EA'} q^E$ with itself. This confirms the first property in (\ref{SD-prop}). 

What we have established is that the space of vectors of the type (\ref{spin-helicity}) is preserved by the Lie bracket, in the sense that the Lie bracket of two such vectors is again a vector of the same type. This immediately shows that the YM bracket (\ref{bracket}) of two such vector fields coincides with their Lie bracket (times two). Indeed, the YM bracket on vector fields that are transverse and that are orthogonal to each other $(\xi_1 \xi_2)=0$ is twice the Lie bracket. This immediately applies to vector fields that are negative helicity polarisation vectors, and shows that on them the YM bracket reduces to the Lie bracket. Lie bracket satisfies the Jacobi identity, which immediately implies that the YM bracket evaluated on negative helicity vector fields also satisfies it. This establishes the colour-kinematics duality for the YM self-dual sector (at 4 points) in a way that is analogous to arguments in \cite{Monteiro:2011pc}, but more directly. 

It is easy to generalise the above argument to an arbitrary number of points. The main property that we need is that the space of vector fields of the form $\xi_{AA'} = q_A \xi_{A'}$, for some spinor $\xi_{A'}$ and with a fixed reference spinor $q_A$ is closed under the Lie bracket. This is essentially the same computation as in (\ref{spinor-bracket}). Thus, Lie bracket of two transverse vector fields of this form is again transverse and is of this form. Then the YM bracket coincides with the Lie bracket, and this happens for an arbitrary number of repeated applications of the YM bracket. Given that the Lie bracket satisfies Jacobi, this establishes the property (\ref{Jacobi}) for the numerators at arbitrary number of gluons. 

Finally, we remark that we have justifiably ignored the 4-valent YM vertex in these considerations, as it is known that this vertex cannot contribute to amplitudes when all helicities are the same. This follows from a simple count of the number of reference spinors $q_A$ that needs to be contracted. 

\section{The Jacobiator of the YM bracket}
\label{sec:jacobiator}

The purpose of this section is to explain the colour-kinematics duality at 4 points. We provide more information on the bracket of vector fields as defined by YM theory Feynman rules, work out the contributions from the gauge-fixing terms (\ref{L3-4-gf}) and interpret these contributions. We also work out the contribution from the 4-valent vertex. The main result of this section is the computation of the Jacobiator of the YM bracket, which measures the failure of the Jacobi identity to be satisfied. We find that the Jacobiator is cancelled by the 4-valent vertex, which explains why the colour-kinematics works at 4 points. We provide a Lie-algebraic explanation of all these facts in the next section. 

\subsection{Metric is not diff-invariant}

We start by noting that the difference between the new bracket (\ref{bracket}) and the Lie bracket (\ref{Lie}) stems from the fact that the flat metric used in the definition (\ref{assoc}) is not diffeomorphism invariant. Indeed, an invariant metric on the Lie algebra satisfies
\be\label{inv-metric}
([Z,X]Y)+(X[Z,Y])=0.
\ee
If this were the case here, which it is not, then we would have all three terms in (\ref{assoc}) equal, and the bracket defined by (\ref{assoc}) was just three times the Lie bracket. The difference thus stems from the non-invariance of the flat metric under the diffeos, as claimed. 

All this can be rephrased as follows. We have the Lie bracket (\ref{Lie}) on vector fields. We build from it (\ref{assoc}) the kinematic factor $v_3(\xi_1,\xi_2,\xi_3)$ that is completely anti-symmetric in its 3 entries, as it should be as it comes from YM Feynman rules, where it is to be multiplied by another anti-symmetric tensor --- the structure constant. We can only build $v_3(\xi_1,\xi_2,\xi_3)$ with the help of the metric. But this metric is not invariant in the sense of (\ref{inv-metric}), and so the kinematic invariant $v_3(\xi_1,\xi_2,\xi_3)$ does not coincide with $([\xi_1,\xi_2]\xi_3)$. Instead, the kinematic factor introduces a new bracket (\ref{cochain}) satisfying (\ref{symmetry}). But this bracket does not satisfy the Jacobi identity. 

\subsection{Gauge-fixing part}

We now perform similar operations with the cubic vertex coming from the gauge-fixing, see (\ref{L3-4-gf}). We get the following (totally anti-symmetric in the 3 vector fields) kinematic factor
\be\label{twist}
v_3^C(\xi_1,\xi_2,\xi_3) = (C\xi_1\xi_2) (k_3\xi_3) +(C\xi_2\xi_3) (k_1\xi_1)+ (C\xi_3\xi_1) (k_2\xi_2).
\ee
Note that this kinematic factor vanishes if all 3 vector fields are transverse $(k_i \xi_i)=0$. Note also that we {\it do not} assume this on-shell condition in what follows. For later purposes, we give another expression for $v_3^C$. Using the momentum conservation, it is easy to see that
\be\label{twist-C}
-v_3^C(\xi_1,\xi_2,\xi_3)= (C[\xi_1,\xi_2]\xi_3) + (C[\xi_3,\xi_1]\xi_2)+(C[\xi_2,\xi_3]\xi_1).
\ee
Note the similar of this with (\ref{v3}). Thus, the anti-symmetric tensor $C$ above plays role analogous to the metric in (\ref{v3}). We will interpret this in terms of twists below. 

This kinematic factor can similarly be viewed as the result of some bracket of vector fields $\xi_1,\xi_2$ contracted with $\xi_3$
\be
[,]_{C}: \Lambda^2 TM \to TM,  \qquad ([\xi_1,\xi_2]_{C} \xi_3):=v_3^C(\xi_1,\xi_2,\xi_3)  
\ee
where
\be\label{bracket-C}
[\xi_1,\xi_2]_{C} = (k_1\xi_1) i_{\xi_2}C-(k_2\xi_2) i_{\xi_1}C  -(C\xi_1\xi_2) (k_1+k_2),
\ee
Here $(i_\xi C)_\mu := \xi^\nu C_{\nu\mu}$ is the interior product. One can also view (\ref{bracket-C}) as a {\it twist} of the bracket (\ref{bracket}) by an element $C$ of $\Lambda^2 T^*M$. 

\subsection{Lie algebra cohomology}

It is possible to understand the expression (\ref{twist}) as an "exact" expression, with an appropriate notion of exactness.
For this purpose, it is useful to introduce some basic notions of Lie algebra cohomology, as this mathematics is very useful to understand what is happening. In our case the Lie algebra is that of vector fields on the manifold. Introduce the notion of {\it cochains} which are multi-linear functions 
\be
f: \Lambda^n TM \to \R.
\ee
Then the {\it coboundary} of an $n$-cochain is the $(n+1)$-cochain $\delta f$ given by
\be
(\delta f)(\xi_1,\ldots,\xi_{n+1}) := \sum_{i<j} (-1)^{i+j} f([\xi_i,\xi_j],\xi_1,\ldots,\hat{\xi}_i,\ldots,\hat{\xi}_j\ldots,\xi_{n+1}).
\ee
The hat indicates that the argument must be omitted. The coboundary operator is nilpotent
\be
\delta^2=0
\ee
and defines the {\it Chevalley-Eilenberg} complex (in our case with values in $\R$). 

\subsection{Twist as a gauge transformation}

Let us now consider a 2-cochain on $\Lambda^2 TM$ constructed using $C\in \Lambda^2 T^*M$
\be
C(\xi_1,\xi_2):=(C\xi_1\xi_2).
\ee
As is easy to check, the kinematic factor $v_3^C(\xi_1,\xi_2,\xi_3)$ that arises as a contribution to the cubic vertex from the gauge-fixing is then just the coboundary of $C$:
\be
v_3^C = \delta C.
\ee
One needs to use the momentum conservation to check this identity. 

\subsection{Coboundary of constant cochains is vanishing on-shell}

As we can see from (\ref{twist}), the coboundary of $C$ from the above example vanishes whenever all 3 vector fields are transverse. This is a general property, as we shall now see. As an illustration, let us compute the coboundary of some 3-cochain $f(\xi_1,\xi_2,\xi_3)$. We have
\be
(\delta f)(\xi_1,\xi_2,\xi_3,\xi_4) = -f([\xi_1,\xi_4],\xi_2,\xi_3)+f([\xi_2,\xi_4],\xi_1,\xi_3)-f([\xi_3,\xi_4],\xi_1,\xi_2)\\ \nonumber
-f([\xi_1,\xi_2],\xi_3,\xi_4)+f([\xi_1,\xi_3],\xi_2,\xi_4)-f([\xi_2,\xi_3],\xi_1,\xi_4).
\ee 
Let us now assume that the cochain $f$ does not contain any derivative operators in it, i.e. is just some completely anti-symmetric rank 3 tensor with vector fields inserted into it. In this case we can use the explicit expression for the Lie bracket and the linearity of the cochain to write
\be
f([\xi_1,\xi_4],\xi_2,\xi_3) = (\xi_1 k_4) f(\xi_4,\xi_2,\xi_3) - (\xi_4 k_1) f(\xi_1,\xi_2,\xi_3).
\ee
It is important that this formula would not be true if the cochain also contains derivative operators, as such derivative operators, when acting on $[\xi_1,\xi_4]$ will need to be replaced by $k_1+k_4$ and the above property would not hold. Keeping this in mind, collecting terms with cochains of the same arguments, and using the momentum conservation, we get
\be\label{coboundary-exact}
(\delta f)(\xi_1,\xi_2,\xi_3,\xi_4) = (\xi_1 k_1) f(\xi_2,\xi_3,\xi_4) -(\xi_2 k_2) f(\xi_1,\xi_3,\xi_4)\\ \nonumber
+(\xi_3 k_3) f(\xi_1,\xi_2,\xi_4)-(\xi_4 k_4) f(\xi_1,\xi_2,\xi_3).
\ee
This vanishes when all 4 vector fields are transverse. Once again, this formula is only true if the cochain $f$ does not contain any derivative operators. E.g. it would not be true for the cochain $v_3$. 

\subsection{Remark}

The expression (\ref{assoc}) for the YM cubic vertex 3-cochain $v_3$ allows one to think about it as being the coboundary of a function on $TM\otimes_S TM$ given by the metric
\be
\eta(\xi_1,\xi_2) := (\xi_1\xi_2).
\ee
Even though the coboundary operation is only defined above on cochains, which $\eta$ is not, we can define
\be\label{eta-coboundary}
(\delta \eta)(\xi_1,\xi_2,\xi_3) := ([\xi_1,\xi_2]\xi_3) + ([\xi_2,\xi_3]\xi_1)+([\xi_3,\xi_1]\xi_2).
\ee
Thus, we have mapped a symmetric function into a completely anti-symmetric one. It is clear that with this definition $\delta \eta = v_3$. But because $\eta$ is not a cochain we have $\delta v_3\not=0$. The fact that in this instance $\delta^2\not=0$ is perhaps the reason not to think of $v_3$ as a coboundary. But we found the analogy too strong to resist mentioning it. 

This remark once again signifies the fact that the only structures that go into the construction of the YM cubic vertex are: (i) the Lie algebra of vector fields with its Lie bracket (\ref{Lie}); (ii) the flat metric $\eta$ that can be used to contract two vector fields. The cubic vertex 3-cochain is then the coboundary (\ref{eta-coboundary}) of the metric, in the sense of Lie algebra cohomology theory. 

\subsection{Properties}

We now list some properties of $[,]_{YM}$. We have for the transverse part of our bracket
\be
([\xi_1,\xi_2]_{YM} (k_1+k_2)) = (\xi_1 k_2)(\xi_2 k_2)-(\xi_1 k_1)(\xi_2 k_1)+ (k_1^2-k_2^2) (\xi_1 \xi_2). 
\ee
This means that when $\xi_1,\xi_2$ are transverse and $k_1^2=k_2^2=0$ (both fields are on-shell), then the result of the bracket of two such vector fields is also transverse. 

It is also interesting to compute the result of the bracket of one vector field with the longitudinal part of the other. Thus, we replace $\xi_1 = k_1$. The result is then
\be\label{k1-x2}
[k_1, \xi_2]_{YM} = -\left(k_2^2 \xi_2 - k_2 (\xi_2 k_2) \right) + \left( (k_1+k_2)^2 \xi_2 - (k_1+k_2) (\xi_2 (k_1+k_2))\right).
\ee
The first term here is a multiple of the orthogonal projection of $\xi_2$ away from $k_2$, and the second term is a multiple of the orthogonal projection of $\xi_2$ away from $k_1+k_2$. In particular, if $\xi_2$ is transverse and on-shell, then (\ref{k1-x2}) is transverse. 

\subsection{Interpreting the quartic vertex}

The quartic part of the Lagrangian (\ref{L3-4}) gives rise to 24 terms. They can be grouped according to how the Lie algebra structure constants contract. There are exactly 3 different contractions, corresponding to the $s,t$ and $u$ channels. We can then put the contribution of the quartic vertex into form (\ref{sum-graphs}) appropriate for colour-kinematics duality. Thus, depending on how the Lie algebra indices contract, we interpret each contribution to the quartic vertex factor as corresponding to either $s,t$ or $u$ channel. We then multiply and divide by the corresponding propagator. This puts the contribution of the quartic vertex into form of a sum over three 3-valent graphs, as in (\ref{sum-graphs}), with certain kinematic factors to be spelled out below. 

Above we have extracted a certain kinematic bracket (\ref{assoc}) from the YM Feynman rules. This was done by stripping off the colour factor of the 3-vertex. When we are to compute the Jacobiator of the YM bracket (\ref{assoc}), we will be adding contributions of different 3-valent graphs, essentially doing the sum as in (\ref{sum-graphs}) but with the colour and the propagator factors stripped off. 

Our task here is to compare the kinematic Jacobiator (to be computed later) with the similar object that can be extracted from the quartic vertex. To extract this we take the quartic vertex factor and multiply and divide each term in it by an appropriate $(k_i+k_j)^2/\im$ according to its colour factor. We then strip both the propagator and the colour factor and add the resulting quantities. It is this object that can be compared to the Jacobiator of the YM vertex (\ref{assoc}).

\subsection{4-cochain}

Taking into account an extra $\im$ from in front of the action, the quartic part of the Lagrangian gives rise to the following structure
\be\nonumber
v_4(\xi_1,\xi_2,\xi_3,\xi_4)=(k_1+k_2)^2 \left( (\xi_1 \xi_3) (\xi_2 \xi_4) - (\xi_2 \xi_3) (\xi_1 \xi_4)\right)\\ \label{v4}
+(k_2+k_3)^2 \left( (\xi_2 \xi_1) (\xi_3 \xi_4) - (\xi_3 \xi_1) (\xi_2 \xi_4)\right)\\ \nonumber
+(k_1+k_3)^2 \left( (\xi_1 \xi_4) (\xi_3 \xi_2) - (\xi_3 \xi_4) (\xi_1 \xi_2)\right).
\ee
It is not hard to check that $v_4$ is actually a 4-cochain, i.e. completely anti-symmetric. 

\subsection{An identity}

Even though the 4-cochain (\ref{v4}) is obtained by adding 3 different contributions of the 4-vertex factor with colours stripped and missing propagators supplied, each term in (\ref{v4}) cannot be interpreted in terms of some product of cubic vertex contributions. However, the whole object (\ref{v4}) can be interpreted in such a way, and this is to play an important role in the next section. 

To establish another expression for (\ref{v4}) we first derive a simple consequence of the momentum conservation. Thus, consider
\be
(k_1+k_2)^2 - (k_2+k_3)^2 = k_1^2 + 2(k_1 k_2) - k_3^2 - 2(k_2 k_3).
\ee
Using the momentum conservation $k_1+k_2+k_3+k_4=0$ we can write the same quantity in a different way as
\be
(k_3+k_4)^2 - (k_1+k_4)^2 = k_3^2 + 2(k_3 k_4) - k_1^2 - 2(k_1 k_4).
\ee
Adding these two expressions we get
\be
(k_1+k_2)^2 - (k_2+k_3)^2 = (k_1 k_2)  - (k_2 k_3)+ (k_3 k_4)  - (k_1 k_4) = ((k_1 -k_3)(k_2-k_4)).
\ee
Using this, we can rewrite (\ref{v4}) as follows
\be\label{v4-identity}
-v_4(\xi_1,\xi_2,\xi_3,\xi_4)= ((k_1 -k_2)(k_3-k_4))(\xi_1 \xi_2)(\xi_3\xi_4) + ((k_3 -k_1)(k_2-k_4))(\xi_3 \xi_1)(\xi_2\xi_4) \\ \nonumber
+((k_1 -k_4)(k_2-k_3))(\xi_1 \xi_4)(\xi_2\xi_3).
\ee
The importance of this identity is that every term here can be interpreted as a product of two cubic vertices of a new type, with each cubic vertex contributing as the bracket
\be
[\xi_1,\xi_2]_* := (k_1-k_2)(\xi_1 \xi_2).
\ee
This fact will be of importance in the next section, when we interpret the objects that we encountered in terms of the Drinfeld double of the group of diffeomorphisms. 

\subsection{Gauge-fixing 4-cochain}

The gauge-fixing term gives, similarly
\be \label{v4-gf}
v_4^C(\xi_1,\xi_2,\xi_3,\xi_4)=(k_1+k_2)^2 (C \xi_1 \xi_2) (C \xi_3 \xi_4) +
(k_2+k_3)^2 (C \xi_2 \xi_3) (C \xi_1 \xi_4) \\ \nonumber + (k_1+k_3)^2  (C \xi_1 \xi_3) (C \xi_4 \xi_2) .
\ee
It can be viewed as a twist of the quartic vertex by an element $C$ of $\Lambda^2 T^*M$. 

\subsection{Jacobiator}

We now perform the computation of the left-hand-side of the would be Jacobi identity for (\ref{bracket}). We call this object the Jacobiator $J:\Lambda^3 TM \to TM$
\be
J[\xi_1,\xi_2,\xi_3]:= [\,[\xi_1,\xi_2]_{YM},\xi_3]_{YM}+[\,[\xi_2,\xi_3]_{YM},\xi_1]_{YM}+[\,[\xi_3,\xi_1]_{YM},\xi_2]_{YM}
\ee
The most convenient way to represent the answer for this quantity is to take its product with some vector field $\xi_4$
\be
J(\xi_1,\xi_2,\xi_3,\xi_4):= (J[\xi_1,\xi_2,\xi_3] \xi_4).
\ee 
The resulting object is a 4-cochain, which can be seen using the symmetry property (\ref{symmetry}) of the bracket. We have
\be\label{J}
J(\xi_1,\xi_2,\xi_3,\xi_4) = ([\xi_1,\xi_2]_{YM} [\xi_3,\xi_4]_{YM}) + ([\xi_2,\xi_3]_{YM} [\xi_1,\xi_4]_{YM})+([\xi_3,\xi_1]_{YM} [\xi_2,\xi_4]_{YM}),
\ee
which is obviously a 4-cochain. 

\subsection{Main result}

We have come to the main result of this section, which is an explicit expression for the Jacobiator of the YM bracket. After a somewhat laborious explicit computation we get
\be\label{Jacobiator}
J = -v_4 + \delta' v_3,
\ee
where $\delta'$ is defined as 
\be\label{delta-prime}
(\delta' f)(\xi_1,\xi_2,\xi_3,\xi_4) := (\xi_1 k_1) f(\xi_2,\xi_3,\xi_4) -(\xi_2 k_2) f(\xi_1,\xi_3,\xi_4)\\ \nonumber
+(\xi_3 k_3) f(\xi_1,\xi_2,\xi_4)-(\xi_4 k_4) f(\xi_1,\xi_2,\xi_3).
\ee
We remind the reader that on 3-cochains that do not have any derivative operators inside, this would coincide with the result of the action of the coboundary operator $\delta$, see (\ref{coboundary-exact}). In general $\delta'$ and $\delta$ are different, and only $\delta$ is nilpotent. The importance of $\delta'$ is that its result is always vanishing on transverse vector fields, as is clear from the above expression. 

The equality (\ref{Jacobiator}) is our main result in this section. In the computation that leads to this result, it is easy to recognise the $v_4$ part on the right-hand-side, in its form (\ref{v4-identity}). It takes more work to rewrite all other terms as $\delta' v_3$. 

The formula (\ref{Jacobiator}) tells us that the failure of the Jacobi identity for $[,]_{YM}$ to be satisfied is "corrected" by the quartic vertex. It cancels the first term on the right-hand-side, and then the Jacobi is satisfied when all 4 external vector fields are transverse. This explains why the colour-kinematics duality works at 4 points.

\subsection{Jacobiator with gauge}

Let us now perform a computation similar to (\ref{Jacobi}) but using the cubic vertex twisted by $C$. Thus, we introduce a new bracket
\be
[,]_{YM}^C: \Lambda^2 TM \to TM,  \qquad ([\xi_1,\xi_2]_{YM}^C \xi_3):=v_3(\xi_1,\xi_2,\xi_3)+v_3^C(\xi_1,\xi_2,\xi_3).  
\ee
Using its symmetry, the Jacobiator is computed as
\be\label{J-C}
J^C(\xi_1,\xi_2,\xi_3,\xi_4) = ([\xi_1,\xi_2]_{YM}^C [\xi_3,\xi_4]_{YM}^C) + ([\xi_2,\xi_3]_{YM}^C [\xi_1,\xi_4]_{YM}^C)+([\xi_3,\xi_1]_{YM}^C [\xi_2,\xi_4]_{YM}^C).
\ee
The computation gives
\be
J^C(\xi_1,\xi_2,\xi_3,\xi_4) = -(v_4+v_4^C) (\xi_1,\xi_2,\xi_3,\xi_4)+(\delta'(v_3+v_3'+v_3^C+v_3^C{}'))(\xi_1,\xi_2,\xi_3,\xi_4)\\ \nonumber
+(\xi_1\xi_2)(C\xi_3\xi_4)(k_1^2-k_2^2) +(\xi_3\xi_4)(C\xi_1\xi_2)(k_3^2-k_4^2)\\ \nonumber
+ (\xi_2\xi_3)(C\xi_1\xi_4)(k_2^2-k_3^2) +(\xi_1\xi_4)(C\xi_2\xi_3)(k_1^2-k_4^2)\\ \nonumber
+(\xi_3\xi_1)(C\xi_2\xi_4)(k_3^2-k_1^2) +(\xi_2\xi_4)(C\xi_3\xi_1)(k_2^2-k_4^2).
\ee
Here
\be
v_3'(\xi_1,\xi_2,\xi_3)=(\xi_1\xi_2)(C\xi_3(k_1-k_2))+(\xi_2\xi_3)(C\xi_1(k_2-k_3))+(\xi_3\xi_1)(C\xi_2(k_3-k_1))
\\ \nonumber
v_3^C{}'(\xi_1,\xi_2,\xi_3)= (C\xi_1\xi_2)(C\xi_3k_3)+(C\xi_2\xi_3)(C\xi_1k_1)+(C\xi_3\xi_1)(C\xi_2k_2).
\ee
We note that both $v_3'$ and $v_3^C{}'$ are obtained from $v_3$ and $v_3^C$ by replacing $\eta^{\mu\nu}$ with $C^{\mu\nu}$ in one of the contractions in each term. 

As before, the quartic vertex precisely cancels the nontrivial part on the right-hand-side of (\ref{J-C}), with all other parts vanishing on-shell. The terms in the first line are zero on-shell being exact, while the terms on lines 2,3,4 are zero when all the 4 momenta are null. 

\section{Drinfeld doubles and twists}
\label{sec:drinfeld}

The purpose of this section is to introduce concepts that later allow us to interpret the result (\ref{Jacobiator}) in terms of a certain twist of the Drinfeld double of the Lie algebra of vector fields. Thus, we review some basics about (classical) Drinfeld doubles. Most of the material in this Section is standard. A source on Lie bi-algebras we found useful is \cite{quasi}.

There is also an unconventional point about our presentation here, as we need to describe the twist of the double by a symmetric tensor. Usually, the literature considers only twists by an anti-symmetric tensor because these do not take one out of the setting of Lie bi-algebras. However, this is insufficient for our purposes, as we will need to twist with a metric, which is a symmetric tensor. The resulting description is rather straightforward, but we were unable to find it in the literature. 

\subsection{Two descriptions}

Drinfeld's construction arose as an axiomatisation of the quantum inverse scattering method. We refer the reader to Drinfeld's original papers \cite{Drinfeld-Hopf}, \cite{Drinfeld-ICM} for more details. 

Drinfeld's construction is quantum, but admits a classical limit. We will only need this classical construction in the present paper. The main structure arising in the classical limit is that of what \cite{Semenov} calls a double Lie algebra. This is a Lie algebra with a second bracket on it also satisfying the Jacobi identity. One can get such double Lie algebras with the help of the classical r-matrices satisfying the classical Yang-Baxter equation.

For our applications, and also in the construction of the Drinfeld double, it is best to think about the second bracket as being defined not on the original vector space, but on its dual. A bracket on the dual space is then equivalent to an operation $\delta: {\mathfrak g}\to \Lambda^2 {\mathfrak g}$ on the original Lie algebra. This motivates the notion of Lie bi-algebras that Drinfeld uses. 

It is convenient to have both pictures: the Lie bi-algebra picture where one works with a single space ${\mathfrak g}$ and two operations on it, and another picture where one works with ${\mathfrak g}\oplus {\mathfrak g}^*$ as well as brackets on both spaces. The later description is the Drinfeld double proper. Some computations are easier in the Lie bi-algebras setting, some in the Drinfeld double picture. We start with the Lie bi-algebra description. 

\subsection{Lie bi-algebras}

This description interprets the bracket on the dual space ${\mathfrak g}^*$ as a 1-cocycle on the original space ${\mathfrak g}$. 

\begin{definition} A Lie bi-algebra is a tuple $({\mathfrak g}, \delta)$, where $\mathfrak g$ is a Lie algebra and $\delta: {\mathfrak g}\to \Lambda^2 {\mathfrak g}$ is a 1-cocycle  such that 
\be\label{co-Jacobi}
{\rm Alt}(\delta\otimes {\rm id})\delta(x) = 0.
\ee
\end{definition}
Here the notion of 1-cocycle is as follows. It is a linear map $\delta: {\mathfrak g}\to \Lambda^2 {\mathfrak g}$ satisfying $\delta([x,y])=[x,\delta(y)]-[y,\delta(x)]$. The operator ${\rm Alt}$ gives a sum over permutations with signs. The property (\ref{co-Jacobi}) is called co-Jacobi identity. The 1-cocycle condition above can be viewed as the condition of compatibility of the bracket $[,]$ on ${\mathfrak g}$ with the co-bracket $\delta$.

\subsection{Twisting}

An important source of Lie bi-algebras is the twisting procedure. Thus, let $r\in \Lambda^2 {\mathfrak g}$ be an object satisfying 
\be\label{r-conditions}
{\rm Alt}(\delta \otimes {\rm id}) r = 0, \qquad [r^{12},r^{13}]+[r^{12},r^{23}]+[r^{13},r^{23}]=0.
\ee
The second of these equations is known as the classical Yang-Baxter equation (CYBE). An object satisfying these properties is referred to as the classical $r$-matrix. 

\begin{lemma} The object
\be\label{delta-twist}
\delta_r(x) = \delta(x) + [x\otimes 1 + 1\otimes x,r] 
\ee
is a 1-cocycle, satisfying (\ref{co-Jacobi}), and thus defines a new Lie bi-algebra. 
\end{lemma}
This Lie bi-algebra is said to be obtained from the original one by twisting via $r$. The 1-cocycle condition is not hard to check. It is harder to check the co-Jacobi identity (\ref{co-Jacobi}), and both properties (\ref{r-conditions}) are necessary for co-Jacobi to be satisfied. 

\bigskip
\noindent {\bf Remark.} It is important that $r$ in the above construction is taken to be an anti-symmetric tensor $r\in \Lambda^2 {\mathfrak g}$. This guarantees that (\ref{delta-twist}) gives an operator mapping $\mathfrak g$ into $\Lambda^2 {\mathfrak g}$. We could also consider symmetric $r$ and change the sign in (\ref{delta-twist}) to continue to get an operator from $\mathfrak g$ to $\Lambda^2 {\mathfrak g}$. However, it is not hard to check that this operator would fail to be a 1-cocycle, i.e. it would not satisfy $\delta[x,y]=[x,\delta(y)]-[y,\delta(x)]$. This is why we can only define twists of Lie bi-algebras by anti-symmetric $r$-matrices. This remark is important for below, because we will need to consider certain twists by symmetric tensors. 

\subsection{Quasi-triangular Drinfeld doubles}

One can start with a trivial Lie bi-algebra for which $\delta=0$, and twist this bi-algebra into a non-trivial one with an $r$-matrix satisfying the CYBE. Lie bi-algebras obtained in this way are called {\it quasi-triangular}. This is an important source of Lie bi-algebras. 

\subsection{The Drinfeld double description}

In this description one interprets the 1-cocycle $\delta$ in terms of a bracket on the dual space ${\mathfrak g}^*$. This is done as follows. 

Let $\mathfrak g$ be a (finite-dimensional) Lie algebra, and $e_i \in {\mathfrak g}$ be some basis. Its Lie bracket can be written as $[e_i, e_j]=C^k_{ij} e_k$, where $C^k_{ij}$ are the structure constants. The Jacobi identity becomes the following quadratic equation satisfied by the structure constants
\be
C^m_{i l} C^l_{jk} + C^m_{j l} C^l_{ki} + C^m_{k l} C^l_{ij} =0.
\ee
Let us now introduce the dual space ${\mathfrak g}^*$ with basis $e^i: e^i(e_j)=\delta^i_j$. We can interpret the operation $\delta: {\mathfrak g}\to \Lambda^2 {\mathfrak g}$ as defining a bracket on the dual space ${\mathfrak g}^*$ via 
\be\label{dual-bracket}
[e^i,e^j]=f^{ij}_k e^k
\ee 
with the structure constants $f^{ij}_k$ being defined by
\be
\delta (e_i) = f^{kl}_i e_k\otimes e_l.
\ee
It is not hard to see that the Jacobi identity for $f$ is then (\ref{co-Jacobi}). It is also not hard to check that the 1-cocycle condition becomes
\be\label{compat}
 f_m^{kl}  C_{ij}^m+f^{km}_i C^l_{jm} -f^{lm}_i C^k_{jm} -  f^{km}_j C^l_{im} +f^{lm}_j C^k_{im}=0,
\ee
which can be viewed as a compatibility between the brackets in ${\mathfrak g}$ and ${\mathfrak g}^*$. 

\begin{definition} The space $D={\mathfrak g}\oplus {\mathfrak g}^*$ with the bracket defined by
\be\label{DD}
[e_i, e_j]=C^k_{ij} e_k, \qquad [e^i,e^j]=f^{ij}_k e^k, \qquad [e_i, e^j]= - C^j_{ik} e^k + f^{jk}_i e_k
\ee
is called the Drinfeld double.
\end{definition}

\begin{lemma} The Drinfeld double is a Lie algebra, i.e. the bracket defined in (\ref{DD}) satisfies the Jacobi identity.
\end{lemma}
A proof is by verification, using both Jacobi identities as well as (\ref{compat}).  

\begin{lemma} The following symmetric tensor  
\be\label{metric}
g= e_i \otimes e^i + e^i\otimes e_i \in D\otimes D,
\ee
with the summation over $i$ implied, is an invariant metric on the Drinfeld double in the sense of (\ref{inv-metric}). 
\end{lemma}
A proof is simple verification. We note that the availability of an invariant metric on $D$, which is not necessarily available in ${\mathfrak g}$, is part of the reason why the double $D$ is an interesting object.

\subsection{Twisting at the level of the Drinfeld double}

To understand the meaning of the twist in (\ref{twist}), let us start with the trivial Drinfeld double with commuting dual space generators $[e^i,e^j]=0$. This trivial double exists for any Lie algebra. Let us then take an arbitrary element $r\in \Lambda^2 {\mathfrak g}$. We can decompose it in the basis
\be
r= r^{ij} e_i \otimes e_j.
\ee
We then define a new basis in the dual space
\be\label{basis-change}
u^i := e^i + r^{ij} e_j.
\ee
We leave the Lie algebra generators unchanged $u_i=e_i$. We have
\be
[u_i, u^j]= [e_i, e^j + r^{jk} e_k] = -C_{ik}^j (u^k -r^{ks} e_s) + r^{jk} C_{ik}^s e_s = - C_{ik}^j u^k + f^{jk}_i e_k,
\ee
where
\be\label{f}
f^{jk}_i = r^{jl}C_{il}^k - r^{kl} C_{il}^j.
\ee
We have used the anti-symmetry of $r^{ij}$ here. We also have that the new generators $u^i$ no longer commute
\be
[u^i, u^j]= [e^i, r^{jk} e_k] + [r^{ik} e_k, e^j] + [r^{ik} e_k, r^{jl} e_l] = r^{jk} C^i_{km} e^m - r^{ik} C^j_{km} e^m + r^{ik} r^{jm} C^s_{km} e_s \\ \nonumber
=r^{jk} C^i_{km}  ( u^m - r^{ms} e_s) -  r^{ik} C^j_{km}( u^m - r^{ms} e_s) + r^{ik} r^{jm} C^s_{km} e_s.
\ee
We now use the classical Yang-Baxter equation (\ref{r-conditions}) to cancel the terms quadratic in $r$ and get
\be
[u^i, u^j]= f^{ij}_k u^k,
\ee
with $f^{ij}_k$ given by (\ref{f}). This coincides with the bracket introduced earlier via (\ref{dual-bracket}), with the co-bracket $\delta$ given by the second term in (\ref{delta-twist}). 

The above construction justifies the earlier introduced terminology of twists. Indeed, the twist (\ref{twist}) is now interpreted as a simple change of basis (\ref{basis-change}) in the Drinfeld double. The new generators $u^i$ are then no longer commuting, but instead form a Lie algebra with structure constants $f^{ij}_k$, provided the twisting $r$-matrix satisfies the CYBE. 

\subsection{Twisting by a symmetric tensor}

As we have already mentioned above, at the level of Lie bi-algebras the twisting (\ref{delta-twist}) can only be done by an anti-symmetric $r$-matrix $r\in \Lambda^2 {\mathfrak g}$. However, at the level of the Drinfeld double $D={\mathfrak g}\oplus{\mathfrak g}^*$ one can consider changes of the basis of the type (\ref{basis-change}) generated by an arbitrary tensor. It is clear that whatever change of basis is performed, the Drinfeld double $D$ remains a Lie algebra. However, what will happen in the process of such twists is that the twisted generators, in general, will not form Lie sub-algebras. Thus, twists by arbitrary tensors take one out of the setting of Lie bi-algebras, while preserving the fact that  $D$ is a Lie algebra. 

Let us illustrate this for the twists that will be of importance for our construction below. Thus, let $\eta^{ij}$ be a symmetric tensor, and $\eta_{ij}$ be its inverse.  We start with a trivial Drinfeld double with commuting dual generators $[e^i,e^j]=0$. We then twist
\be\label{twist-u}
u_i := e_i + \eta_{ij} e^j, \qquad u^i :=e^i - \eta^{ij} e_j.
\ee
The inverse of these transformations is
\be
e_i = \frac{1}{2}(u_i - \eta_{ij} u^j), \qquad e^i = \frac{1}{2}(u^i + \eta^{ij} u_j).
\ee
Let us compute, for future reference, the invariant metric (\ref{metric}) on the Drinfeld double on the new generators. We can write the metric (\ref{metric}) as 
\be
\la e_i, e^j\ra=\delta_i^j,
\ee
with all other products being zero. This gives
\be\label{DD-metric}
\la u_i, u_j \ra = 2\eta_{ij}, \quad \la u^i , u^j \ra= - 2\eta^{ij}, \qquad \la u_i, u^j \ra=0.
\ee
Thus, with respect to the invariant metric on $D$ the two subspaces spanned by $u_i, u^i$ are now orthogonal. Note that now, once embedded into the Drinfeld double, the metric $\eta_{ij}$, which is of course not an invariant metric on the original Lie algebra, becomes the invariant metric on generators $u_i, u_j$.

For future convenience, let us introduce a notation for the parts of $D$ that are spanned by $u_i, u^i$
\be
{\mathfrak u} := {\rm Span}(u_i), \qquad {\mathfrak u}^\perp := {\rm Span}(u^i).
\ee
The second notation here is justified, because $u^i$ are indeed orthogonal to $u_i$ with respect to the invariant metric on $D$. Thus, we can write
\be
D = {\mathfrak u}\oplus {\mathfrak u}^\perp.
\ee

\subsection{A new bracket}

We now compute the brackets between the new generators. We have
\be\label{uu}
[u_i, u_j]=[e_i + \eta_{ik} e^k, e_j + \eta_{jl} e^l] = \left( C^m_{ij} e_m - \eta_{jl} C^l_{im} e^m + \eta_{ik} C^k_{jm} e^m\right) \\ \nonumber
= \frac{1}{2} C^m_{ij} (u_m -\eta_{mn} u^n) - \frac{1}{2} \eta_{jl} C^l_{im} (u^m+ \eta^{mn} u_n) + \frac{1}{2} \eta_{ik} C^k_{jm}(u^m+ \eta^{mn} u_n) \\ \nonumber
=  {}^\eta C^k_{ij} u_k -  A_{ijk} u^k,
\ee
where
\be\label{C-eta}
{}^\eta C^k_{ij} = \frac{1}{2}\left( C^k_{ij} +\eta^{km} C^l_{mi} \eta_{jl} + \eta^{km} C^l_{jm} \eta_{il}\right),
\quad
A_{ijk} = \frac{1}{2} \left( C_{ij}^m \eta_{mk} + C^m_{ik} \eta_{jm}- C^m_{jk}\eta_{im} \right).
\ee
Projecting the right-hand-side on the ${\mathfrak u}$ part of $D$ we get a new bracket
\be\label{uu-eta}
[u_i, u_j]_\eta := [u_i,u_j]\Big|_{{\mathfrak u}} = {}^\eta C^k_{ij}  u_k.
\ee

\subsection{A different representation}

It is useful to write the above bracket in a different form. Thus, let us introduce vectors $a= a^i u_i$. We then have two possible ways to commute such objects. In the first of these, we remember that $u_i$ is a vector in the Drinfeld double given by (\ref{twist-u}), and use the Drinfeld double bracket $[,]$. In the second, we first project the vector $a$ onto ${\mathfrak g}$ and then take the bracket of the result. Thus, let us define
\be\label{uu-prime}
[a_1,a_2]':=[a_1|_{\mathfrak g}, a_2|_{\mathfrak g}] = a_1^i a_2^j C_{ij}^k e_k.
\ee
We can then rewrite the bracket (\ref{uu-eta}) in terms of this bracket. The bracket (\ref{uu-eta}) is simply the projection of the bracket in the double onto the ${\mathfrak u}$ part. This can be selected by taking the product with another element of ${\mathfrak u}$. Thus, we have
\be\label{eta-bracket}
\la [a_1,a_2], a_3\ra = \la[a_1,a_2]_\eta ,a_3\ra = \la[a_1,a_2]' ,a_3\ra+\la [a_2,a_3]' ,a_1\ra +\la [a_3,a_1]' ,a_2\ra,
\ee
where to write the second equality we have used the definition (\ref{uu-prime}), as well as (\ref{C-eta}). We recognise exactly the same structure (\ref{cochain}), (\ref{assoc}) as one defining what we called the YM bracket above. 

\subsection{Jacobi identity after the twist}

We have found that the bracket (\ref{eta-bracket}) arises in the process of twisting (\ref{twist-u}) the Drinfeld double. The information that goes into the twist is that of a symmetric tensor $\eta^{ij}$ and its inverse. This is also the information that is necessary to define the bracket (\ref{eta-bracket}). There is now no mystery as to why the bracket (\ref{eta-bracket}) does not satisfy the Jacobi identity, and the Drinfeld double construction also shows that a certain version of the Jacobi identity is still satisfied.

To see what happens, let us compute the Jacobiator of elements $a_{1,2,3}$ with $a_1= a_1^i u_i$, etc., and project on the fourth such element. The result is of course zero, because the original bracket satisfies Jacobi. We have
\be
0= \la [[a_1,a_2],a_3],a_4\ra +\la[[a_2,a_3],a_1],a_4\ra+\la[[a_3,a_1],a_2],a_4\ra \\ \nonumber
= \la[a_1,a_2],[a_3,a_4]\ra + \la[a_2,a_3],[a_1,a_4]\ra+\la[a_3,a_1],[a_2,a_4]\ra.
\ee
We have used the invariance of the metric to obtain the second expression. As we see from (\ref{uu}), the result of each bracket here contains part in the subspace spanned by $u_i$, as well as in the subspace spanned by $u^i$. We now use the fact that the $[,]_\eta$ bracket is just the projection of the full bracket on the space spanned by $u_i$, as well as orthogonality of the spaces spanned by $u_i$ and $u^i$. There is also a relative minus sign in the Drinfeld double metric (\ref{DD-metric}). Dividing by 2, we get the following identity
\be
{}^\eta C^{m}_{ij} \, {}^\eta C^{n}_{kl} \eta_{mn} +{}^\eta C^{m}_{jk} \, {}^\eta C^{n}_{il} \eta_{mn} +{}^\eta C^{m}_{ki} \, {}^\eta C^{n}_{jl} \eta_{mn} \\ \nonumber =  A_{ijm} A_{kln} \eta^{mn} + A_{jkm} A_{iln} \eta^{mn} + A_{kim} A_{jln} \eta^{mn}.
\ee
The first line here is just the Jacobiator cochain (\ref{J}), while the second line is the measure of the failure of the Jacobi identity for the bracket (\ref{eta-bracket}) to be satisfied. 

It is worth representing this identity in graphical terms. Let straight lines represent the space spanned by $u_i$, and straight lines with a dot in the middle represent the projection onto the space spanned by $u^i$. The above identity then takes the form of the usual Jacobi identity (\ref{Jacobi}), but one has to keep in mind that the intermediate channel can be either in the space spanned by $u_i$ or in the orthogonal space. One also has to keep in mind a relative minus sign of the metric in the ${\mathfrak u}^\perp$ part of $D$. So, we get
\fmfcmd{%
style_def dot expr p =
cdraw p;
filldraw fullcircle scaled 5 shifted point length(p)/2 of p
enddef;}
\bigskip
\begin{align}   \label{Jacobi-mod}
0= \qquad  \parbox{20mm}{\begin{fmfgraph*}(20,20) 
  \fmftop{i1,i2}
  \fmfbottom{o1,o2}
  \fmflabel{1}{i1}
   \fmflabel{2}{i2}
    \fmflabel{4}{o1}
     \fmflabel{3}{o2}
\fmf{plain}{i1,v1}
\fmf{plain}{i2,v1}
\fmf{plain}{v1,v2}
  \fmf{plain}{v2,o1}
  \fmf{plain}{v2,o2}
   \end{fmfgraph*}}\,\,   \quad +   \quad \parbox{20mm}{\begin{fmfgraph*}(20,20) 
  \fmftop{i1,i2}
  \fmfbottom{o1,o2}
  \fmflabel{2}{i1}
   \fmflabel{3}{i2}
    \fmflabel{4}{o1}
     \fmflabel{1}{o2}
\fmf{plain}{i1,v1}
\fmf{plain}{i2,v1}
\fmf{plain}{v1,v2}
  \fmf{plain}{v2,o1}
  \fmf{plain}{v2,o2}
   \end{fmfgraph*}}\,\, \quad+ \quad \,\, \parbox{20mm}{\begin{fmfgraph*}(20,20) 
  \fmftop{i1,i2}
  \fmfbottom{o1,o2}
  \fmflabel{3}{i1}
   \fmflabel{1}{i2}
    \fmflabel{4}{o1}
     \fmflabel{2}{o2}
\fmf{plain}{i1,v1}
\fmf{plain}{i2,v1}
\fmf{plain}{v1,v2}
  \fmf{plain}{v2,o1}
  \fmf{plain}{v2,o2}
   \end{fmfgraph*}}\,\, \\ \nonumber \vspace{0.5in} \\ \nonumber + \quad  \,\, \parbox{20mm}{\begin{fmfgraph*}(20,20) 
  \fmftop{i1,i2}
  \fmfbottom{o1,o2}
  \fmflabel{1}{i1}
   \fmflabel{2}{i2}
    \fmflabel{4}{o1}
     \fmflabel{3}{o2}
\fmf{plain}{i1,v1}
\fmf{plain}{i2,v1}
\fmf{dot}{v1,v2}
  \fmf{plain}{v2,o1}
  \fmf{plain}{v2,o2}
   \end{fmfgraph*}}\,\,   \quad +   \quad \parbox{20mm}{\begin{fmfgraph*}(20,20) 
  \fmftop{i1,i2}
  \fmfbottom{o1,o2}
  \fmflabel{2}{i1}
   \fmflabel{3}{i2}
    \fmflabel{4}{o1}
     \fmflabel{1}{o2}
\fmf{plain}{i1,v1}
\fmf{plain}{i2,v1}
\fmf{dot}{v1,v2}
  \fmf{plain}{v2,o1}
  \fmf{plain}{v2,o2}
   \end{fmfgraph*}}\,\, \quad+ \quad \,\, \parbox{20mm}{\begin{fmfgraph*}(20,20) 
  \fmftop{i1,i2}
  \fmfbottom{o1,o2}
  \fmflabel{3}{i1}
   \fmflabel{1}{i2}
    \fmflabel{4}{o1}
     \fmflabel{2}{o2}
\fmf{plain}{i1,v1}
\fmf{plain}{i2,v1}
\fmf{dot}{v1,v2}
  \fmf{plain}{v2,o1}
  \fmf{plain}{v2,o2}
   \end{fmfgraph*}}
\end{align}
\bigskip

\noindent In the next section we will compare this identity to (\ref{Jacobiator}) that we have derived for the YM bracket. 

\section{The Drinfeld double of the Lie algebra of vector fields}
\label{sec:DD}

This section interprets the result (\ref{Jacobiator}) in terms of the Drinfeld double of the Lie algebra of diffeomorphisms. We will see that the proper way to understand the appearance of the YM bracket, as well as the quartic vertex and its role in making the Jacobi identity in the form (\ref{Jacobiator}) satisfied is by a twist of the Drinfeld double. We have already identified the required twist in the previous section, see (\ref{twist-u}). In this section we continue to work at 4 points, postponing the situation at an arbitrary number of gluons till the next section.

We would like to point out from the outset that the Drinfeld double that we start with is the trivial one, where the bracket on the dual space is zero $f^{ij}_k=0$, see previous section for details. Such a double exists for any Lie algebra. 

\subsection{The Drinfeld double}

Let us first spell out what the Drinfeld double construction gives us for the case of the Lie algebra of vector fields. As we recall from the previous section, the Drinfeld double construction gives a Lie bracket on ${\mathfrak g}\oplus{\mathfrak g}^*$ from a Lie bracket on ${\mathfrak g}$. To compute the bracket between ${\mathfrak g}$ and ${\mathfrak g}^*$ we note that the last expression in (\ref{DD}) is equivalent to the following definition. Let $a_1,a_2\in {\mathfrak g}$ and $b\in {\mathfrak g}^*$. Then $[a_1,b]\in {\mathfrak g}^*$ is defined so that $\la[a_1,b],a_2\ra=-\la b, [a_1,a_2]\ra$. Note that this is simply the requirement that the metric is invariant. With this way of defining the bracket between ${\mathfrak g}$ and ${\mathfrak g}^*$, it is easy to see that we get the following expression for the Drinfeld double bracket between vector fields and one forms
\be
[\xi, \eta] = {\cal L}_\xi \eta + \eta (\partial \xi), \qquad \xi\in TM, \quad \eta\in T^*M.
\ee
We would like to stress that this bracket explicitly depends on the metric, namely, in $(\partial\xi)$ in the last term. In momentum space terms we have
\be
[\xi,\eta] = (\xi(k_1+k_2)) \eta + (\xi\eta) k_1.
\ee
We can also write the full bracket on the Drinfeld double 
\be
D = TM \oplus T^*M.
\ee
A general element of $D$ is thus a sum of a vector field and a one-form $\xi + \eta$. We have
\be\label{DD-bracket}
[\xi_1+\eta_1, \xi_2+\eta_2]= [\xi_1,\xi_2]+ {\cal L}_{\xi_1} \eta_2 - {\cal L}_{\xi_2} \eta_1 + \eta_2 (\partial \xi_1) - \eta_1 (\partial \xi_2),
\ee
where $[\xi_1,\xi_2]$ is the usual Lie bracket of vector fields. 
We have explicitly checked that this bracket satisfies Jacobi identity in the sense that
\be
[[\xi_1,\xi_2],\eta]+[[\xi_2,\eta],\xi_1]+[[\eta,\xi_1],\xi_2]=0.
\ee
The part of Jacobi identity with two or three instances of a one-form are trivially satisfied because one-forms commute. We note that the Drinfeld double bracket (\ref{DD-bracket}) is similar to the so-called Courant bracket put to prominent use in \cite{Hitchin}. Both brackets are on the sum of the tangent and cotangent bundles. The difference is in the last two terms, and in the fact that the Courant bracket does {\it not} satisfy Jacobi identity. 

\subsection{The twist}

The twist that we need to consider to get to the YM bracket is essentially the same as what is used to encode the metric in Hitchin's generalised geometry, see \cite{Hitchin-05}, page 3. 

Thus, we twist the Drinfeld double in the sense of (\ref{twist-u}). That is, with every vector field $\xi\in TM$ we associate an element of the Drinfeld double $\xi + \xi^*$, where $\xi^* \in T^* M$ is the one-form obtained by lowering the index of the vector $\xi$ using the flat metric. On elements of this form, the invariant metric on the Drinfeld double is just (a multiple of) the flat metric $\eta$. 

The Drinfeld double bracket of two such elements is not longer an element of this form, but it can be projected on such elements. This defines a new bracket on vector fields, which we shall refer to as the YM bracket, as it coincides with (\ref{bracket}), as we will now check. We have
\be
([\xi_1,\xi_2]_{YM}\xi_3)\equiv  \la [\xi_1+\xi_1^*,\xi_2+\xi_2^*],(\xi_3+\xi_3^*)\ra = \la [\xi_1,\xi_2],\xi_3^*\ra+ \la [\xi_1,\xi_2^*],\xi_3\ra +\la [\xi_1^*,\xi_2],\xi_3\ra \\ \nonumber
= \la[\xi_1,\xi_2],\xi_3^*\ra + \la[\xi_3,\xi_1],\xi_2^*\ra +\la[\xi_2,\xi_3],\xi_1^*\ra = ([\xi_1,\xi_2]\xi_3) + ([\xi_2,\xi_3]\xi_1) +([\xi_3,\xi_1]\xi_2) ,
\ee 
where we have used the invariance (\ref{inv-metric}) of the Drinfeld double metric to get the first expression in the second line. Here $\la,\ra$ is the invariant metric on the Drinfeld double and $(\xi_1,\xi_2)=\eta_{\mu\nu} \xi_1^\mu \xi_2^\nu$ is the metric pairing of vector fields. We recognise here the YM bracket as defined via (\ref{assoc}), (\ref{cochain}). 

\subsection{The modified Jacobi}

Thus, we have explained the YM bracket (\ref{bracket}) as arising via the simple twist $\xi\to \xi+\xi^*$ of the Drinfeld double with the bracket (\ref{DD-bracket}). As we also know from considerations of the previous section, the failure of the Jacobi for the YM bracket to be satisfied is explained by the fact that in $[[\xi_1,\xi_2]_{YM},\xi_3]_{YM}$ we project on the ${\mathfrak u}$ part of the Drinfeld double $D$ before computing the second bracket. The missing terms contain the terms in the ${\mathfrak u}^\perp$ part. 

Let us compute these terms. Thus, we consider
\be
([\xi_1+\xi_1^*,\xi_2+\xi_2^*](\xi_3-\xi_3^*)) = -([\xi_1,\xi_2]\xi_3^*) + ([\xi_3,\xi_1]\xi_2^*) +([\xi_2,\xi_3]\xi_1^*).
\ee
Writing all terms here in the momentum space we get
\be
([\xi_1+\xi_1^*,\xi_2+\xi_2^*](\xi_3-\xi_3^*)):= ([\xi_1,\xi_2]_{YM}^\perp \xi_3),
\ee
where
\be
[\xi_1,\xi_2]_{YM}^\perp = (\xi_1\xi_2)(k_1-k_2) + (\xi_1 k_1) \xi_2 - (\xi_2 k_2) \xi_1.
\ee

Let us now compute the terms contributing to the Jacobi identity. We have
\be
([\xi_1,\xi_2]_{YM}^\perp [\xi_3,\xi_4]_{YM}^\perp) = ((k_1-k_2)(k_3-k_4))(\xi_1\xi_2)(\xi_3\xi_4) \\ \nonumber
+(\xi_1\xi_2)( (k_1-k_2)( (\xi_3 k_3)\xi_4-(\xi_4 k_4)\xi_3)) + (\xi_3\xi_4)( (k_3-k_4)( (\xi_1 k_1)\xi_2-(\xi_2 k_2)\xi_1)) 
\\ \nonumber
+\left(( (\xi_1 k_1)\xi_2-(\xi_2 k_2)\xi_1) ( (\xi_3 k_3)\xi_4-(\xi_4 k_4)\xi_3)\right).
\ee
The first of the terms on the right-hand-side is what appears in (\ref{v4-identity}). All other terms are vanishing on-shell. Collecting them requires some work, but the end result is
\be\label{dd-4-points}
([\xi_1,\xi_2]_{YM}^\perp [\xi_3,\xi_4]_{YM}^\perp) +([\xi_3,\xi_1]_{YM}^\perp [\xi_2,\xi_4]_{YM}^\perp) +([\xi_1,\xi_4]_{YM}^\perp [\xi_2,\xi_3]_{YM}^\perp) 
= - v_4 + \delta' v_3,
\ee
where $v_4$ is as in (\ref{v4-identity}) and $\delta' v_3$ is given by (\ref{delta-prime}). 

We now take into account the fact that the Drinfeld double metric (\ref{DD-metric}) on elements of the form $\xi-\xi^*$ is negative definite. There is thus an extra minus sign in comparing the contributions from ${\mathfrak u}$ and ${\mathfrak u}^\perp$. Thus, we see that the Drinfled double Jacobi identity takes the form
\be
([\xi_1,\xi_2]_{YM} [\xi_3,\xi_4]_{YM}) +([\xi_3,\xi_1]_{YM} [\xi_2,\xi_4]_{YM}) +([\xi_1,\xi_4]_{YM} [\xi_2,\xi_3]_{YM}) \\ \nonumber
=([\xi_1,\xi_2]_{YM}^\perp [\xi_3,\xi_4]_{YM}^\perp) +([\xi_3,\xi_1]_{YM}^\perp [\xi_2,\xi_4]_{YM}^\perp) +([\xi_1,\xi_4]_{YM}^\perp [\xi_2,\xi_3]_{YM}^\perp) ,
\ee
and that this coincides with (\ref{Jacobiator}). This completes the interpretation of (\ref{Jacobiator}) that we discovered by a computation following from the YM Feynman rules, as the Jacobi identity in the Drinfled double of the Lie algebra of vector fields. 

\subsection{Gauge-fixing freedom as twist}

Here we interpret the gauge-fixing freedom (\ref{gauge-fix}) parametrised by an anti-symmetric tensor $C^{\mu\nu}$ as another type of twist of the Drinfeld double. This is the twist by an anti-symmetric tensor, and is thus of the type that was already considered in the previous section and that by itself does not take one out of the setting of Lie bi-algebras. Here we combine this twist with the already considered twist by a symmetric tensor. The full twist is a twist of the Drinfeld double and does not preserve the Lie bi-algebra setting. 

Thus, let us consider the tensor 
\be\label{X}
X_{\mu\nu} := \eta_{\mu\nu} + C_{\mu\nu},
\ee
which no longer has any specific symmetry property, as a sum of the symmetric metric and anti-symmetric $C$. We can use this tensor to lower the indices of vector fields
\be
X: TM\to T^* M, \qquad \xi^\mu \to \xi^X_\mu:= \xi^\nu X_{\nu\mu} .
\ee
We can then consider elements of the Drinfeld double of the form
\be\label{xi-X}
\xi +\xi^X.
\ee
The orthogonal complement in $D$ to elements of this form are elements
\be
 \xi - \xi^{X^T},
\ee
where $X^T$ is the transpose $X^T_{\mu\nu}=X_{\nu\mu}$. 

Let us consider the Drinfeld double bracket of two elements of the form (\ref{xi-X}), projected again on an element of this form. We have
\be
\la[\xi_1+\xi_1^X,\xi_2+\xi_2^X],\xi_3+\xi_3^X\ra=\la [\xi_1,\xi_2],\xi_3^X \ra + \la[\xi_1,\xi_2^X],\xi_3\ra + \la [\xi_1^X,\xi_2],\xi_3\ra \\ \nonumber
=\la [\xi_1,\xi_2],\xi_3^X \ra + \la[\xi_3,\xi_1],\xi_2^X \ra + \la [\xi_2,\xi_3],\xi_1^X\ra = v_3(\xi_1,\xi_2,\xi_3)+v_3^C(\xi_1,\xi_2,\xi_3),
\ee
where $v_3^C$ is given by (\ref{twist-C}). Thus, we see that the gauge-fixing freedom (\ref{gauge-fix}) just corresponds to the availability of more general twists (\ref{xi-X}), where the twisting tensor is a sum of a symmetric metric and an anti-symmetric tensor $C$. 

\section{Five points}

Here we apply the structure discovered at four points to the problem of constructing a colour-kinematics dual expression for amplitudes at 5 points. The non-triviality of this requirement will already be seen at this level, and generalisations to higher points, if at all possible, will follow the pattern to be identified in this case. 

\subsection{Feynman diagrams at five points}

To be as explicit as possible, we start by drawing all (groups of) Feynman diagrams that can arise at 5 points. As in the case of four points considered above, we would like to read diagrams as maps from a set of gluons to the final gluon state. We take the gluon number 5 to represent the final state. The topology of 3-valent diagrams is such that the gluon number 5 sits either at the end of a diagram, or in the middle, see pictures below. 

Let us first draw diagrams where gluon 5 is at the end. This can always be drawn as being at the bottom of the diagram. We get 4 groups of 3 diagrams, each group consisting of diagrams like

\bigskip
\begin{align}\label{5-b}
 \parbox{20mm}{\begin{fmfgraph*}(20,20) 
  \fmftop{i1,i2}
  \fmfright{i3}
  \fmfbottom{i4,i5}
  \fmflabel{1}{i1}
   \fmflabel{2}{i2}
    \fmflabel{3}{i3}
     \fmflabel{5}{i4}
      \fmflabel{4}{i5}
\fmf{wiggly}{i1,v1}
\fmf{wiggly}{i2,v1}
\fmf{wiggly}{v1,v2}
  \fmf{wiggly,tension=0}{v2,i3}
  \fmf{wiggly}{v2,v3}
  \fmf{wiggly}{v3,i4}
   \fmf{wiggly}{v3,i5}
   \end{fmfgraph*}}\,\,   \quad \quad +  \quad  \parbox{20mm}{\begin{fmfgraph*}(20,20) 
  \fmftop{i1,i2}
  \fmfright{i3}
  \fmfbottom{i4,i5}
  \fmflabel{2}{i1}
   \fmflabel{3}{i2}
    \fmflabel{1}{i3}
     \fmflabel{5}{i4}
      \fmflabel{4}{i5}
\fmf{wiggly}{i1,v1}
\fmf{wiggly}{i2,v1}
\fmf{wiggly}{v1,v2}
  \fmf{wiggly,tension=0}{v2,i3}
  \fmf{wiggly}{v2,v3}
  \fmf{wiggly}{v3,i4}
   \fmf{wiggly}{v3,i5}
   \end{fmfgraph*}}\,\, \quad \quad+  \quad \parbox{20mm}{\begin{fmfgraph*}(20,20) 
 \fmftop{i1,i2}
  \fmfright{i3}
  \fmfbottom{i4,i5}
  \fmflabel{3}{i1}
   \fmflabel{1}{i2}
    \fmflabel{2}{i3}
     \fmflabel{5}{i4}
      \fmflabel{4}{i5}
\fmf{wiggly}{i1,v1}
\fmf{wiggly}{i2,v1}
\fmf{wiggly}{v1,v2}
  \fmf{wiggly,tension=0}{v2,i3}
  \fmf{wiggly}{v2,v3}
  \fmf{wiggly}{v3,i4}
   \fmf{wiggly}{v3,i5}
   \end{fmfgraph*}}
\end{align}
\bigskip

\noindent plus $3\times 3$ more diagrams like this, with gluon 4 in the bottom right being replaced by $1,2,3$ respectively. 

Then, there are 3 diagrams where the gluon 5 is in the middle of the diagram

\bigskip
\begin{align}\label{5-m}
 \parbox{20mm}{\begin{fmfgraph*}(20,20) 
  \fmfleft{i1,i2}
  \fmfright{i3,i4}
  \fmfbottom{i5}
  \fmflabel{1}{i1}
   \fmflabel{2}{i2}
    \fmflabel{4}{i3}
     \fmflabel{3}{i4}
      \fmflabel{5}{i5}
\fmf{wiggly}{i1,v1}
\fmf{wiggly}{i2,v1}
\fmf{wiggly}{v1,v2}
  \fmf{wiggly,tension=0}{v2,i5}
  \fmf{wiggly}{v2,v3}
  \fmf{wiggly}{v3,i3}
   \fmf{wiggly}{v3,i4}
   \end{fmfgraph*}}\,\,   \quad \quad +  \quad  \parbox{20mm}{\begin{fmfgraph*}(20,20) 
 \fmfleft{i1,i2}
  \fmfright{i3,i4}
  \fmfbottom{i5}
  \fmflabel{3}{i1}
   \fmflabel{1}{i2}
    \fmflabel{4}{i3}
     \fmflabel{2}{i4}
      \fmflabel{5}{i5}
\fmf{wiggly}{i1,v1}
\fmf{wiggly}{i2,v1}
\fmf{wiggly}{v1,v2}
  \fmf{wiggly,tension=0}{v2,i5}
  \fmf{wiggly}{v2,v3}
  \fmf{wiggly}{v3,i3}
   \fmf{wiggly}{v3,i4}
   \end{fmfgraph*}}\,\, \quad \quad+  \quad \parbox{20mm}{\begin{fmfgraph*}(20,20) 
\fmfleft{i1,i2}
  \fmfright{i3,i4}
  \fmfbottom{i5}
  \fmflabel{1}{i1}
   \fmflabel{4}{i2}
    \fmflabel{3}{i3}
     \fmflabel{2}{i4}
      \fmflabel{5}{i5}
\fmf{wiggly}{i1,v1}
\fmf{wiggly}{i2,v1}
\fmf{wiggly}{v1,v2}
  \fmf{wiggly,tension=0}{v2,i5}
  \fmf{wiggly}{v2,v3}
  \fmf{wiggly}{v3,i3}
   \fmf{wiggly}{v3,i4}
   \end{fmfgraph*}}
\end{align}
\bigskip

We then draw Feynman diagrams containing a 4-valent vertex. There are 4 diagrams where the gluon 5 is not inserted into the 4-valent vertex

\bigskip
\begin{align}
 \parbox{20mm}{\begin{fmfgraph*}(20,20) 
  \fmftop{i1,i2,i3}
  \fmfright{i4}
  \fmfbottom{i5}
  \fmflabel{1}{i1}
   \fmflabel{2}{i2}
    \fmflabel{3}{i3}
     \fmflabel{4}{i4}
      \fmflabel{5}{i5}
\fmf{wiggly}{i1,v1}
\fmf{wiggly}{i2,v1}
\fmf{wiggly}{i3,v1}
 \fmf{wiggly}{v1,v2}
  \fmf{wiggly,tension=0}{v2,i4}
  \fmf{wiggly}{v2,i5}
    \end{fmfgraph*}}\,\,   \quad \quad +  \quad  \parbox{20mm}{\begin{fmfgraph*}(20,20) 
  \fmftop{i1,i2,i3}
  \fmfright{i4}
  \fmfbottom{i5}
  \fmflabel{1}{i1}
   \fmflabel{4}{i2}
    \fmflabel{2}{i3}
     \fmflabel{3}{i4}
      \fmflabel{5}{i5}
\fmf{wiggly}{i1,v1}
\fmf{wiggly}{i2,v1}
\fmf{wiggly}{i3,v1}
 \fmf{wiggly}{v1,v2}
  \fmf{wiggly,tension=0}{v2,i4}
  \fmf{wiggly}{v2,i5}
   \end{fmfgraph*}}\,\, \quad \quad+  \quad \parbox{20mm}{\begin{fmfgraph*}(20,20) 
 \fmftop{i1,i2,i3}
  \fmfright{i4}
  \fmfbottom{i5}
  \fmflabel{1}{i1}
   \fmflabel{3}{i2}
    \fmflabel{4}{i3}
     \fmflabel{2}{i4}
      \fmflabel{5}{i5}
\fmf{wiggly}{i1,v1}
\fmf{wiggly}{i2,v1}
\fmf{wiggly}{i3,v1}
 \fmf{wiggly}{v1,v2}
  \fmf{wiggly,tension=0}{v2,i4}
  \fmf{wiggly}{v2,i5}
   \end{fmfgraph*}}\,\, \quad \quad+  \quad \parbox{20mm}{\begin{fmfgraph*}(20,20) 
 \fmftop{i1,i2,i3}
  \fmfright{i4}
  \fmfbottom{i5}
  \fmflabel{3}{i1}
   \fmflabel{2}{i2}
    \fmflabel{4}{i3}
     \fmflabel{1}{i4}
      \fmflabel{5}{i5}
\fmf{wiggly}{i1,v1}
\fmf{wiggly}{i2,v1}
\fmf{wiggly}{i3,v1}
 \fmf{wiggly}{v1,v2}
  \fmf{wiggly,tension=0}{v2,i4}
  \fmf{wiggly}{v2,i5}
   \end{fmfgraph*}}
\end{align}
\bigskip

Finally, there are 6 diagrams where the gluon 5 participates in the 4-valent vertex

\bigskip
\begin{align}
 \parbox{20mm}{\begin{fmfgraph*}(20,20) 
  \fmftop{i1,i2}
  \fmfbottom{i5,i4,i3}
  \fmflabel{1}{i1}
   \fmflabel{2}{i2}
    \fmflabel{3}{i3}
     \fmflabel{4}{i4}
      \fmflabel{5}{i5}
\fmf{wiggly}{i1,v1}
\fmf{wiggly}{i2,v1}
 \fmf{wiggly}{v1,v2}
   \fmf{wiggly}{v2,i3}
     \fmf{wiggly}{v2,i4}
   \fmf{wiggly}{v2,i5}
    \end{fmfgraph*}}\,\,   \quad \quad +  \quad  \parbox{20mm}{\begin{fmfgraph*}(20,20) 
  \fmftop{i1,i2}
  \fmfbottom{i5,i4,i3}
  \fmflabel{3}{i1}
   \fmflabel{1}{i2}
    \fmflabel{2}{i3}
     \fmflabel{4}{i4}
      \fmflabel{5}{i5}
\fmf{wiggly}{i1,v1}
\fmf{wiggly}{i2,v1}
 \fmf{wiggly}{v1,v2}
   \fmf{wiggly}{v2,i3}
     \fmf{wiggly}{v2,i4}
   \fmf{wiggly}{v2,i5}
   \end{fmfgraph*}}\,\, \quad \quad+  \quad \parbox{20mm}{\begin{fmfgraph*}(20,20) 
 \fmftop{i1,i2}
  \fmfbottom{i5,i4,i3}
  \fmflabel{2}{i1}
   \fmflabel{3}{i2}
    \fmflabel{1}{i3}
     \fmflabel{4}{i4}
      \fmflabel{5}{i5}
\fmf{wiggly}{i1,v1}
\fmf{wiggly}{i2,v1}
 \fmf{wiggly}{v1,v2}
   \fmf{wiggly}{v2,i3}
     \fmf{wiggly}{v2,i4}
   \fmf{wiggly}{v2,i5}
   \end{fmfgraph*}}\\ \nonumber \\ \nonumber \vspace{0.5in} \\ \nonumber
   +  \quad \parbox{20mm}{\begin{fmfgraph*}(20,20) 
  \fmftop{i1,i2}
  \fmfbottom{i5,i4,i3}
  \fmflabel{1}{i1}
   \fmflabel{4}{i2}
    \fmflabel{2}{i3}
     \fmflabel{3}{i4}
      \fmflabel{5}{i5}
\fmf{wiggly}{i1,v1}
\fmf{wiggly}{i2,v1}
 \fmf{wiggly}{v1,v2}
   \fmf{wiggly}{v2,i3}
     \fmf{wiggly}{v2,i4}
   \fmf{wiggly}{v2,i5}
   \end{fmfgraph*}}\,\, \quad \quad+  \quad \parbox{20mm}{\begin{fmfgraph*}(20,20) 
 \fmftop{i1,i2}
  \fmfbottom{i5,i4,i3}
  \fmflabel{2}{i1}
   \fmflabel{4}{i2}
    \fmflabel{3}{i3}
     \fmflabel{1}{i4}
      \fmflabel{5}{i5}
\fmf{wiggly}{i1,v1}
\fmf{wiggly}{i2,v1}
 \fmf{wiggly}{v1,v2}
   \fmf{wiggly}{v2,i3}
     \fmf{wiggly}{v2,i4}
   \fmf{wiggly}{v2,i5}
   \end{fmfgraph*}}\,\, \quad \quad+  \quad \parbox{20mm}{\begin{fmfgraph*}(20,20) 
 \fmftop{i1,i2}
  \fmfbottom{i5,i4,i3}
  \fmflabel{3}{i1}
   \fmflabel{4}{i2}
    \fmflabel{1}{i3}
     \fmflabel{2}{i4}
      \fmflabel{5}{i5}
\fmf{wiggly}{i1,v1}
\fmf{wiggly}{i2,v1}
 \fmf{wiggly}{v1,v2}
   \fmf{wiggly}{v2,i3}
     \fmf{wiggly}{v2,i4}
   \fmf{wiggly}{v2,i5}
   \end{fmfgraph*}}
\end{align}
\bigskip

\subsection{Numerators}

As is well-known, see e.g. \cite{BjerrumBohr:2010zs}, there are 15 different numerators at 5 points. Each of these numerators corresponds to one of the 3-valent diagrams (\ref{5-b}), (\ref{5-m}). Then can be labelled by specifying one of the pairs at the sides of the diagram, together with the middle leg. Thus, e.g. the numerator corresponding to the first diagram in (\ref{5-b}) can be labelled as $n_{12}^3$. We read the diagram from the top, and so we have $n_{45}^3=-n_{12}^3$. 

Diagrams that are 3-valent correspond to a unique numerator. However, each numerator also receives a contribution from diagrams with 4-valent vertex. For example, the numerator $n_{12}^3$ is given by
\be
n_{12}^3 =  s_{12} s_{45} \parbox{20mm}{\begin{fmfgraph*}(20,20) 
  \fmftop{i1,i2}
  \fmfright{i3}
  \fmfbottom{i4,i5}
  \fmflabel{1}{i1}
   \fmflabel{2}{i2}
    \fmflabel{3}{i3}
     \fmflabel{5}{i4}
      \fmflabel{4}{i5}
\fmf{wiggly}{i1,v1}
\fmf{wiggly}{i2,v1}
\fmf{wiggly}{v1,v2}
  \fmf{wiggly,tension=0}{v2,i3}
  \fmf{wiggly}{v2,v3}
  \fmf{wiggly}{v3,i4}
   \fmf{wiggly}{v3,i5}
   \end{fmfgraph*}}\,\,   \quad \quad +  \quad  s_{12} s_{45} \parbox{20mm}{\begin{fmfgraph*}(20,20) 
  \fmftop{i1,i2,i3}
  \fmfright{i4}
  \fmfbottom{i5}
  \fmflabel{1}{i1}
   \fmflabel{2}{i2}
    \fmflabel{3}{i3}
     \fmflabel{4}{i4}
      \fmflabel{5}{i5}
\fmf{wiggly}{i1,v1}
\fmf{wiggly}{i2,v1}
\fmf{wiggly}{i3,v1}
 \fmf{wiggly}{v1,v2}
  \fmf{wiggly,tension=0}{v2,i4}
  \fmf{wiggly}{v2,i5}
    \end{fmfgraph*}}\,\,   \quad \quad +  s_{12} s_{45} \quad  \parbox{20mm}{\begin{fmfgraph*}(20,20) 
  \fmftop{i1,i2}
  \fmfbottom{i5,i4,i3}
  \fmflabel{1}{i1}
   \fmflabel{2}{i2}
    \fmflabel{3}{i3}
     \fmflabel{4}{i4}
      \fmflabel{5}{i5}
\fmf{wiggly}{i1,v1}
\fmf{wiggly}{i2,v1}
 \fmf{wiggly}{v1,v2}
   \fmf{wiggly}{v2,i3}
     \fmf{wiggly}{v2,i4}
   \fmf{wiggly}{v2,i5}
    \end{fmfgraph*}}
    \ee
    \bigskip
    
\noindent Here $s_{ij}=(k_i+k_j)^2$ is the (inverse of) the propagator. It is understood that each diagram is with the colour structure constants stripped off. It is also understood that only part of the second and third diagram with the colour contraction as in the first diagram contribute. In the first diagram the factor $s_{12} s_{45}$ removes the propagators, while in the second and third diagrams one propagator (for the single intermediate line) is removed in each, and a factor of $s$ is introduced.

\subsection{Identities at 4 points}

To help understand the situation at 5 points, let us introduce a graphical representation of the identities we discovered at 4 point. Thus, we have seen that the colour- and propagator- stripped 3-valent Feynman graph have the interpretation of a successive application of the YM bracket. We continue to denote this bracket by a cubic vertex sourced by straight lines. Thus, we have
\be
s_{12} \parbox{20mm}{\begin{fmfgraph*}(20,20) 
  \fmftop{i1,i2}
  \fmfbottom{i4,i3}
  \fmflabel{1}{i1}
   \fmflabel{2}{i2}
    \fmflabel{3}{i3}
     \fmflabel{4}{i4}
     \fmf{wiggly}{i1,v1}
\fmf{wiggly}{i2,v1}
 \fmf{wiggly}{v1,v2}
   \fmf{wiggly}{v2,i3}
     \fmf{wiggly}{v2,i4}
    \end{fmfgraph*}} \qquad = \qquad \parbox{20mm}{\begin{fmfgraph*}(20,20) 
  \fmftop{i1,i2}
  \fmfbottom{i4,i3}
  \fmflabel{1}{i1}
   \fmflabel{2}{i2}
    \fmflabel{3}{i3}
     \fmflabel{4}{i4}
     \fmf{plain}{i1,v1}
\fmf{plain}{i2,v1}
 \fmf{plain}{v1,v2}
   \fmf{plain}{v2,i3}
     \fmf{plain}{v2,i4}
    \end{fmfgraph*}} \qquad \equiv ([\xi_1,\xi_2]_{YM} [\xi_3,\xi_4]_{YM}).
\ee
\bigskip

\noindent Here on the left we have a colour-stripped Feynman graph, which is also multiplied by $s_{12}$ to remove the propagator. On the right we have a quantity that has the Drinfeld double interpretation, as explained in the previous section. 

We have also seen in (\ref{v4-identity}) that the sum of colour-stripped contributions from the 4-valent vertex, with missing propagators introduced can be seen as a sum over 3-valent graphs, modulo terms that vanish on transverse vector fields
\begin{align}\label{identity}
v_4\equiv s_{12} \parbox{20mm}{\begin{fmfgraph*}(20,20) 
  \fmftop{i1,i2}
  \fmfbottom{i4,i3}
  \fmflabel{1}{i1}
   \fmflabel{2}{i2}
    \fmflabel{3}{i3}
     \fmflabel{4}{i4}
     \fmf{wiggly}{i1,v1}
\fmf{wiggly}{i2,v1}
    \fmf{wiggly}{v1,i3}
     \fmf{wiggly}{v1,i4}
    \end{fmfgraph*}}  \qquad + \quad s_{23} \parbox{20mm}{\begin{fmfgraph*}(20,20) 
  \fmftop{i1,i2}
  \fmfbottom{i4,i3}
  \fmflabel{2}{i1}
   \fmflabel{3}{i2}
    \fmflabel{1}{i3}
     \fmflabel{4}{i4}
     \fmf{wiggly}{i1,v1}
\fmf{wiggly}{i2,v1}
    \fmf{wiggly}{v1,i3}
     \fmf{wiggly}{v1,i4}
    \end{fmfgraph*}} \qquad + \quad s_{13} \parbox{20mm}{\begin{fmfgraph*}(20,20) 
  \fmftop{i1,i2}
  \fmfbottom{i4,i3}
  \fmflabel{3}{i1}
   \fmflabel{1}{i2}
    \fmflabel{2}{i3}
     \fmflabel{4}{i4}
     \fmf{wiggly}{i1,v1}
\fmf{wiggly}{i2,v1}
    \fmf{wiggly}{v1,i3}
     \fmf{wiggly}{v1,i4}
    \end{fmfgraph*}}\,\, \\ \nonumber \vspace{0.5in} \\ \nonumber =  \quad  \,\, \parbox{20mm}{\begin{fmfgraph*}(20,20) 
  \fmftop{i1,i2}
  \fmfbottom{o1,o2}
  \fmflabel{1}{i1}
   \fmflabel{2}{i2}
    \fmflabel{4}{o1}
     \fmflabel{3}{o2}
\fmf{plain}{i1,v1}
\fmf{plain}{i2,v1}
\fmf{dot}{v1,v2}
  \fmf{plain}{v2,o1}
  \fmf{plain}{v2,o2}
   \end{fmfgraph*}}\,\,   \quad +   \quad \,\,\parbox{20mm}{\begin{fmfgraph*}(20,20) 
  \fmftop{i1,i2}
  \fmfbottom{o1,o2}
  \fmflabel{2}{i1}
   \fmflabel{3}{i2}
    \fmflabel{4}{o1}
     \fmflabel{1}{o2}
\fmf{plain}{i1,v1}
\fmf{plain}{i2,v1}
\fmf{dot}{v1,v2}
  \fmf{plain}{v2,o1}
  \fmf{plain}{v2,o2}
   \end{fmfgraph*}}\,\, \quad+ \quad \,\, \parbox{20mm}{\begin{fmfgraph*}(20,20) 
  \fmftop{i1,i2}
  \fmfbottom{o1,o2}
  \fmflabel{3}{i1}
   \fmflabel{1}{i2}
    \fmflabel{4}{o1}
     \fmflabel{2}{o2}
\fmf{plain}{i1,v1}
\fmf{plain}{i2,v1}
\fmf{dot}{v1,v2}
  \fmf{plain}{v2,o1}
  \fmf{plain}{v2,o2}
   \end{fmfgraph*}}\,\, \quad + \quad {\rm exact\,\,\, terms.}
\end{align}
\bigskip

\noindent Here "exact terms" stands for the terms denoted by $\delta' v_3$ in (\ref{dd-4-points}). We have introduced the following graphical notation
\be\label{black-dot-interpr}
\parbox{20mm}{\begin{fmfgraph*}(20,20) 
  \fmftop{i1,i2}
  \fmfbottom{o1,o2}
  \fmflabel{1}{i1}
   \fmflabel{2}{i2}
    \fmflabel{4}{o1}
     \fmflabel{3}{o2}
\fmf{plain}{i1,v1}
\fmf{plain}{i2,v1}
\fmf{dot}{v1,v2}
  \fmf{plain}{v2,o1}
  \fmf{plain}{v2,o2}
   \end{fmfgraph*}} \qquad \equiv - ([\xi_1,\xi_2]^\perp_{YM} [\xi_3,\xi_4]^\perp_{YM}).
\ee
\bigskip

\noindent  Again, this quantity has a Drinfeld double interpretation, as we have explored in the previous section. The minus sign on the right-hand-side is due to the metric on the ${\mathfrak u}^\perp$ part of $D$ being negative definite. 

It is important to emphasise that the identity (\ref{identity}) holds only for the sum of these terms. Thus, we have the following important failure of two quantities to be equal
\be\label{property}
s_{12} \parbox{20mm}{\begin{fmfgraph*}(20,20) 
  \fmftop{i1,i2}
  \fmfbottom{i4,i3}
  \fmflabel{1}{i1}
   \fmflabel{2}{i2}
    \fmflabel{3}{i3}
     \fmflabel{4}{i4}
     \fmf{wiggly}{i1,v1}
\fmf{wiggly}{i2,v1}
    \fmf{wiggly}{v1,i3}
     \fmf{wiggly}{v1,i4}
    \end{fmfgraph*}}  \qquad \not=  \qquad \parbox{20mm}{\begin{fmfgraph*}(20,20) 
  \fmftop{i1,i2}
  \fmfbottom{o1,o2}
  \fmflabel{1}{i1}
   \fmflabel{2}{i2}
    \fmflabel{4}{o1}
     \fmflabel{3}{o2}
\fmf{plain}{i1,v1}
\fmf{plain}{i2,v1}
\fmf{dot}{v1,v2}
  \fmf{plain}{v2,o1}
  \fmf{plain}{v2,o2}
   \end{fmfgraph*}} \,\, \quad + \quad {\rm e.\,\,\, t.}
   \ee
   \bigskip
   
\noindent Here ${\rm e. t.}$ stands for "exact terms". This failure of an equality to hold is the principal reason why we cannot produce explicit colour-kinematics satisfying numerators beyond 4 points. 

\subsection{Sum of numerators at 5 points}

We now check what the above identities at 4 points imply for the sum of numerators at 5 points. The colour-kinematics satisfying numerators at 4 points should satisfy 3-term identities. For instance, one must have $n_{12}^3+n_{23}^1+n_{31}^2=0$. Let us compute this sum using the graphical representation of all the quantities. We have
\begin{align} \label{n-sum}
n_{12}^3 +n_{23}^1+n_{31}^2 =  s_{12} s_{45} \parbox{20mm}{\begin{fmfgraph*}(20,20) 
  \fmftop{i1,i2}
  \fmfright{i3}
  \fmfbottom{i4,i5}
  \fmflabel{1}{i1}
   \fmflabel{2}{i2}
    \fmflabel{3}{i3}
     \fmflabel{5}{i4}
      \fmflabel{4}{i5}
\fmf{wiggly}{i1,v1}
\fmf{wiggly}{i2,v1}
\fmf{wiggly}{v1,v2}
  \fmf{wiggly,tension=0}{v2,i3}
  \fmf{wiggly}{v2,v3}
  \fmf{wiggly}{v3,i4}
   \fmf{wiggly}{v3,i5}
   \end{fmfgraph*}}\,\,   \quad \quad +  \quad  s_{12} s_{45} \parbox{20mm}{\begin{fmfgraph*}(20,20) 
  \fmftop{i1,i2,i3}
  \fmfright{i4}
  \fmfbottom{i5}
  \fmflabel{1}{i1}
   \fmflabel{2}{i2}
    \fmflabel{3}{i3}
     \fmflabel{4}{i4}
      \fmflabel{5}{i5}
\fmf{wiggly}{i1,v1}
\fmf{wiggly}{i2,v1}
\fmf{wiggly}{i3,v1}
 \fmf{wiggly}{v1,v2}
  \fmf{wiggly,tension=0}{v2,i4}
  \fmf{wiggly}{v2,i5}
    \end{fmfgraph*}}\,\,   \quad \quad +  s_{12} s_{45} \quad  \parbox{20mm}{\begin{fmfgraph*}(20,20) 
  \fmftop{i1,i2}
  \fmfbottom{i5,i4,i3}
  \fmflabel{1}{i1}
   \fmflabel{2}{i2}
    \fmflabel{3}{i3}
     \fmflabel{4}{i4}
      \fmflabel{5}{i5}
\fmf{wiggly}{i1,v1}
\fmf{wiggly}{i2,v1}
 \fmf{wiggly}{v1,v2}
   \fmf{wiggly}{v2,i3}
     \fmf{wiggly}{v2,i4}
   \fmf{wiggly}{v2,i5}
    \end{fmfgraph*}} \\ \nonumber  \\ \nonumber  \\ \nonumber 
    s_{23} s_{45} \parbox{20mm}{\begin{fmfgraph*}(20,20) 
  \fmftop{i1,i2}
  \fmfright{i3}
  \fmfbottom{i4,i5}
  \fmflabel{2}{i1}
   \fmflabel{3}{i2}
    \fmflabel{1}{i3}
     \fmflabel{5}{i4}
      \fmflabel{4}{i5}
\fmf{wiggly}{i1,v1}
\fmf{wiggly}{i2,v1}
\fmf{wiggly}{v1,v2}
  \fmf{wiggly,tension=0}{v2,i3}
  \fmf{wiggly}{v2,v3}
  \fmf{wiggly}{v3,i4}
   \fmf{wiggly}{v3,i5}
   \end{fmfgraph*}}\,\,   \quad \quad +  \quad  s_{23} s_{45} \parbox{20mm}{\begin{fmfgraph*}(20,20) 
  \fmftop{i1,i2,i3}
  \fmfright{i4}
  \fmfbottom{i5}
  \fmflabel{1}{i1}
   \fmflabel{2}{i2}
    \fmflabel{3}{i3}
     \fmflabel{4}{i4}
      \fmflabel{5}{i5}
\fmf{wiggly}{i1,v1}
\fmf{wiggly}{i2,v1}
\fmf{wiggly}{i3,v1}
 \fmf{wiggly}{v1,v2}
  \fmf{wiggly,tension=0}{v2,i4}
  \fmf{wiggly}{v2,i5}
    \end{fmfgraph*}}\,\,   \quad \quad +  s_{23} s_{45} \quad  \parbox{20mm}{\begin{fmfgraph*}(20,20) 
  \fmftop{i1,i2}
  \fmfbottom{i5,i4,i3}
  \fmflabel{2}{i1}
   \fmflabel{3}{i2}
    \fmflabel{1}{i3}
     \fmflabel{4}{i4}
      \fmflabel{5}{i5}
\fmf{wiggly}{i1,v1}
\fmf{wiggly}{i2,v1}
 \fmf{wiggly}{v1,v2}
   \fmf{wiggly}{v2,i3}
     \fmf{wiggly}{v2,i4}
   \fmf{wiggly}{v2,i5}
    \end{fmfgraph*}} \\ \nonumber  \\ \nonumber  \\ \nonumber 
    s_{13} s_{45} \parbox{20mm}{\begin{fmfgraph*}(20,20) 
  \fmftop{i1,i2}
  \fmfright{i3}
  \fmfbottom{i4,i5}
  \fmflabel{3}{i1}
   \fmflabel{1}{i2}
    \fmflabel{2}{i3}
     \fmflabel{5}{i4}
      \fmflabel{4}{i5}
\fmf{wiggly}{i1,v1}
\fmf{wiggly}{i2,v1}
\fmf{wiggly}{v1,v2}
  \fmf{wiggly,tension=0}{v2,i3}
  \fmf{wiggly}{v2,v3}
  \fmf{wiggly}{v3,i4}
   \fmf{wiggly}{v3,i5}
   \end{fmfgraph*}}\,\,   \quad \quad +  \quad  s_{13} s_{45} \parbox{20mm}{\begin{fmfgraph*}(20,20) 
  \fmftop{i1,i2,i3}
  \fmfright{i4}
  \fmfbottom{i5}
  \fmflabel{1}{i1}
   \fmflabel{2}{i2}
    \fmflabel{3}{i3}
     \fmflabel{4}{i4}
      \fmflabel{5}{i5}
\fmf{wiggly}{i1,v1}
\fmf{wiggly}{i2,v1}
\fmf{wiggly}{i3,v1}
 \fmf{wiggly}{v1,v2}
  \fmf{wiggly,tension=0}{v2,i4}
  \fmf{wiggly}{v2,i5}
    \end{fmfgraph*}}\,\,   \quad \quad +  s_{13} s_{45} \quad  \parbox{20mm}{\begin{fmfgraph*}(20,20) 
  \fmftop{i1,i2}
  \fmfbottom{i5,i4,i3}
  \fmflabel{3}{i1}
   \fmflabel{1}{i2}
    \fmflabel{2}{i3}
     \fmflabel{4}{i4}
      \fmflabel{5}{i5}
\fmf{wiggly}{i1,v1}
\fmf{wiggly}{i2,v1}
 \fmf{wiggly}{v1,v2}
   \fmf{wiggly}{v2,i3}
     \fmf{wiggly}{v2,i4}
   \fmf{wiggly}{v2,i5}
    \end{fmfgraph*}}     
    \end{align}
    \bigskip
    
We now proceed to replacing the objects here with quantities that have a Lie-algebraic interpretation. For the sum of cubic graphs this is immediate
\begin{align}\label{sum-3}
\parbox{20mm}{\begin{fmfgraph*}(20,20) 
  \fmftop{i1,i2}
  \fmfright{i3}
  \fmfbottom{i4,i5}
  \fmflabel{1}{i1}
   \fmflabel{2}{i2}
    \fmflabel{3}{i3}
     \fmflabel{5}{i4}
      \fmflabel{4}{i5}
\fmf{wiggly}{i1,v1}
\fmf{wiggly}{i2,v1}
\fmf{wiggly}{v1,v2}
  \fmf{wiggly,tension=0}{v2,i3}
  \fmf{wiggly}{v2,v3}
  \fmf{wiggly}{v3,i4}
   \fmf{wiggly}{v3,i5}
   \end{fmfgraph*}}\,\,   \quad \quad + s_{23} s_{45} \parbox{20mm}{\begin{fmfgraph*}(20,20) 
  \fmftop{i1,i2}
  \fmfright{i3}
  \fmfbottom{i4,i5}
  \fmflabel{2}{i1}
   \fmflabel{3}{i2}
    \fmflabel{1}{i3}
     \fmflabel{5}{i4}
      \fmflabel{4}{i5}
\fmf{wiggly}{i1,v1}
\fmf{wiggly}{i2,v1}
\fmf{wiggly}{v1,v2}
  \fmf{wiggly,tension=0}{v2,i3}
  \fmf{wiggly}{v2,v3}
  \fmf{wiggly}{v3,i4}
   \fmf{wiggly}{v3,i5}
   \end{fmfgraph*}}\,\,   \quad \quad + s_{13} s_{45} \parbox{20mm}{\begin{fmfgraph*}(20,20) 
  \fmftop{i1,i2}
  \fmfright{i3}
  \fmfbottom{i4,i5}
  \fmflabel{3}{i1}
   \fmflabel{1}{i2}
    \fmflabel{2}{i3}
     \fmflabel{5}{i4}
      \fmflabel{4}{i5}
\fmf{wiggly}{i1,v1}
\fmf{wiggly}{i2,v1}
\fmf{wiggly}{v1,v2}
  \fmf{wiggly,tension=0}{v2,i3}
  \fmf{wiggly}{v2,v3}
  \fmf{wiggly}{v3,i4}
   \fmf{wiggly}{v3,i5}
   \end{fmfgraph*}} \\ \nonumber \\ \nonumber
 =\qquad  \parbox{20mm}{\begin{fmfgraph*}(20,20) 
  \fmftop{i1,i2}
  \fmfright{i3}
  \fmfbottom{i4,i5}
  \fmflabel{1}{i1}
   \fmflabel{2}{i2}
    \fmflabel{3}{i3}
     \fmflabel{5}{i4}
      \fmflabel{4}{i5}
\fmf{plain}{i1,v1}
\fmf{plain}{i2,v1}
\fmf{plain}{v1,v2}
  \fmf{plain,tension=0}{v2,i3}
  \fmf{plain}{v2,v3}
  \fmf{plain}{v3,i4}
   \fmf{plain}{v3,i5}
   \end{fmfgraph*}}\,\,   \quad \quad +  \parbox{20mm}{\begin{fmfgraph*}(20,20) 
  \fmftop{i1,i2}
  \fmfright{i3}
  \fmfbottom{i4,i5}
  \fmflabel{2}{i1}
   \fmflabel{3}{i2}
    \fmflabel{1}{i3}
     \fmflabel{5}{i4}
      \fmflabel{4}{i5}
\fmf{plain}{i1,v1}
\fmf{plain}{i2,v1}
\fmf{plain}{v1,v2}
  \fmf{plain,tension=0}{v2,i3}
  \fmf{plain}{v2,v3}
  \fmf{plain}{v3,i4}
   \fmf{plain}{v3,i5}
   \end{fmfgraph*}}\,\,   \quad \quad +  \parbox{20mm}{\begin{fmfgraph*}(20,20) 
  \fmftop{i1,i2}
  \fmfright{i3}
  \fmfbottom{i4,i5}
  \fmflabel{3}{i1}
   \fmflabel{1}{i2}
    \fmflabel{2}{i3}
     \fmflabel{5}{i4}
      \fmflabel{4}{i5}
\fmf{plain}{i1,v1}
\fmf{plain}{i2,v1}
\fmf{plain}{v1,v2}
  \fmf{plain,tension=0}{v2,i3}
  \fmf{plain}{v2,v3}
  \fmf{plain}{v3,i4}
   \fmf{plain}{v3,i5}
   \end{fmfgraph*}} 
   \end{align}
   \bigskip
   
   \noindent For the sum of second terms in each line in (\ref{n-sum}) we use (\ref{identity}) to get
   
   \bigskip
   \begin{align}\label{sum-4}
   s_{12} s_{45} \parbox{20mm}{\begin{fmfgraph*}(20,20) 
  \fmftop{i1,i2,i3}
  \fmfright{i4}
  \fmfbottom{i5}
  \fmflabel{1}{i1}
   \fmflabel{2}{i2}
    \fmflabel{3}{i3}
     \fmflabel{4}{i4}
      \fmflabel{5}{i5}
\fmf{wiggly}{i1,v1}
\fmf{wiggly}{i2,v1}
\fmf{wiggly}{i3,v1}
 \fmf{wiggly}{v1,v2}
  \fmf{wiggly,tension=0}{v2,i4}
  \fmf{wiggly}{v2,i5}
    \end{fmfgraph*}}\,\,   \quad \quad +\quad  s_{23} s_{45} \parbox{20mm}{\begin{fmfgraph*}(20,20) 
  \fmftop{i1,i2,i3}
  \fmfright{i4}
  \fmfbottom{i5}
  \fmflabel{1}{i1}
   \fmflabel{2}{i2}
    \fmflabel{3}{i3}
     \fmflabel{4}{i4}
      \fmflabel{5}{i5}
\fmf{wiggly}{i1,v1}
\fmf{wiggly}{i2,v1}
\fmf{wiggly}{i3,v1}
 \fmf{wiggly}{v1,v2}
  \fmf{wiggly,tension=0}{v2,i4}
  \fmf{wiggly}{v2,i5}
    \end{fmfgraph*}}\,\,   \quad \quad +\quad  s_{13} s_{45} \parbox{20mm}{\begin{fmfgraph*}(20,20) 
  \fmftop{i1,i2,i3}
  \fmfright{i4}
  \fmfbottom{i5}
  \fmflabel{1}{i1}
   \fmflabel{2}{i2}
    \fmflabel{3}{i3}
     \fmflabel{4}{i4}
      \fmflabel{5}{i5}
\fmf{wiggly}{i1,v1}
\fmf{wiggly}{i2,v1}
\fmf{wiggly}{i3,v1}
 \fmf{wiggly}{v1,v2}
  \fmf{wiggly,tension=0}{v2,i4}
  \fmf{wiggly}{v2,i5}
    \end{fmfgraph*}} \\ \nonumber \\ \nonumber
 =\qquad  \parbox{20mm}{\begin{fmfgraph*}(20,20) 
  \fmftop{i1,i2}
  \fmfright{i3}
  \fmfbottom{i4,i5}
  \fmflabel{1}{i1}
   \fmflabel{2}{i2}
    \fmflabel{3}{i3}
     \fmflabel{5}{i4}
      \fmflabel{4}{i5}
\fmf{plain}{i1,v1}
\fmf{plain}{i2,v1}
\fmf{dot}{v1,v2}
  \fmf{plain,tension=0}{v2,i3}
  \fmf{plain}{v2,v3}
  \fmf{plain}{v3,i4}
   \fmf{plain}{v3,i5}
   \end{fmfgraph*}}\,\,   \quad \quad +  \parbox{20mm}{\begin{fmfgraph*}(20,20) 
  \fmftop{i1,i2}
  \fmfright{i3}
  \fmfbottom{i4,i5}
  \fmflabel{2}{i1}
   \fmflabel{3}{i2}
    \fmflabel{1}{i3}
     \fmflabel{5}{i4}
      \fmflabel{4}{i5}
\fmf{plain}{i1,v1}
\fmf{plain}{i2,v1}
\fmf{dot}{v1,v2}
  \fmf{plain,tension=0}{v2,i3}
  \fmf{plain}{v2,v3}
  \fmf{plain}{v3,i4}
   \fmf{plain}{v3,i5}
   \end{fmfgraph*}}\,\,   \quad \quad +  \parbox{20mm}{\begin{fmfgraph*}(20,20) 
  \fmftop{i1,i2}
  \fmfright{i3}
  \fmfbottom{i4,i5}
  \fmflabel{3}{i1}
   \fmflabel{1}{i2}
    \fmflabel{2}{i3}
     \fmflabel{5}{i4}
      \fmflabel{4}{i5}
\fmf{plain}{i1,v1}
\fmf{plain}{i2,v1}
\fmf{dot}{v1,v2}
  \fmf{plain,tension=0}{v2,i3}
  \fmf{plain}{v2,v3}
  \fmf{plain}{v3,i4}
   \fmf{plain}{v3,i5}
   \end{fmfgraph*}} \,\, \quad + \quad {\rm e.\,\,\, t.}
    \end{align}
\bigskip

\noindent Unfortunately, we cannot perform a similar operation on the sum of the last term in each line in (\ref{n-sum}). In all these terms, after the intermediate line propagator is removed, what is left is the same kinematic factor $s_{45}$ in front. 

Even in the absence of the equality sign in (\ref{property}) we can further simplify things by noting that the sum of (\ref{sum-3}) and (\ref{sum-4}) vanishes by the already established property at 4 points. Indeed, since in this sum we sum over both ways ${\mathfrak u},{\mathfrak u}^\perp$ that the intermediate top line can be projected, this sum equals to the sum of 3 Lie algebra diagrams with no projection on the top line. Thus, this is just the Jacobiator of 3 vector fields $\xi_1+\xi_1^*,\xi_2+\xi_2^*,\xi_3+\xi_3^*$ times the bracket of vector fields $\xi_4+\xi_4^*,\xi_5+\xi_5^*$, with the projector on ${\mathfrak u}$ inserted
\be
\la [\xi_1+\xi_1^*,\xi_2+\xi_2^*,\xi_3+\xi_3^*]\Big|_{\mathfrak u} [\xi_4+\xi_4^*,\xi_5+\xi_5^*] \ra =0,
\ee
because of the Jacobi identity satisfied by the bracket of the Drinfeld double. Here 
\be
[\xi_1,\xi_2,\xi_3] := [[\xi_1,\xi_2],\xi_3]+[[\xi_2,\xi_3],\xi_1]+[[\xi_3,\xi_1],\xi_2].
\ee

Thus, we see that the sum of the kinematic numerators in (\ref{n-sum}) equals to just the sum of the last terms in each line. Each of these terms is proportional to $s_{45}$, and so the sum of the kinematics numerators $n_{12}^3+n_{23}^1+n_{31}^2$ is also proportional to $s_{45}$. This is all we can conclude from the Feynman rules, as well as using the Jacobi identity established at 4 points. 

\subsection{If there was an equal sign in (\ref{property})}

We have thus confirmed that the kinematic numerators $n_{ij}^k$ produced by the Feynman rules do not satisfy the colour-kinematic duality, which is a known fact. However, our Lie-algebraic interpretation suggests the following way to correct this. 

Thus, let us assume that there is some other way of writing the Feynman rules (and perhaps some other twist of the Drinfeld double) in which there is an equal sign in (\ref{property}). We make some remarks on how this could be possible in the last section. We could then use this property individually for the last terms in each line in (\ref{n-sum}). Then the sum of these terms would be
\begin{align}\label{third-terms}
s_{12} s_{45} \quad  \parbox{20mm}{\begin{fmfgraph*}(20,20) 
  \fmftop{i1,i2}
  \fmfbottom{i5,i4,i3}
  \fmflabel{1}{i1}
   \fmflabel{2}{i2}
    \fmflabel{3}{i3}
     \fmflabel{4}{i4}
      \fmflabel{5}{i5}
\fmf{wiggly}{i1,v1}
\fmf{wiggly}{i2,v1}
 \fmf{wiggly}{v1,v2}
   \fmf{wiggly}{v2,i3}
     \fmf{wiggly}{v2,i4}
   \fmf{wiggly}{v2,i5}
    \end{fmfgraph*}}\qquad +  s_{23} s_{45} \quad  \parbox{20mm}{\begin{fmfgraph*}(20,20) 
  \fmftop{i1,i2}
  \fmfbottom{i5,i4,i3}
  \fmflabel{2}{i1}
   \fmflabel{3}{i2}
    \fmflabel{1}{i3}
     \fmflabel{4}{i4}
      \fmflabel{5}{i5}
\fmf{wiggly}{i1,v1}
\fmf{wiggly}{i2,v1}
 \fmf{wiggly}{v1,v2}
   \fmf{wiggly}{v2,i3}
     \fmf{wiggly}{v2,i4}
   \fmf{wiggly}{v2,i5}
    \end{fmfgraph*}} \qquad +  s_{13} s_{45} \quad  \parbox{20mm}{\begin{fmfgraph*}(20,20) 
  \fmftop{i1,i2}
  \fmfbottom{i5,i4,i3}
  \fmflabel{3}{i1}
   \fmflabel{1}{i2}
    \fmflabel{2}{i3}
     \fmflabel{4}{i4}
      \fmflabel{5}{i5}
\fmf{wiggly}{i1,v1}
\fmf{wiggly}{i2,v1}
 \fmf{wiggly}{v1,v2}
   \fmf{wiggly}{v2,i3}
     \fmf{wiggly}{v2,i4}
   \fmf{wiggly}{v2,i5}
    \end{fmfgraph*}}     \\ \nonumber \\ \nonumber
   =\qquad  \parbox{20mm}{\begin{fmfgraph*}(20,20) 
  \fmftop{i1,i2}
  \fmfright{i3}
  \fmfbottom{i4,i5}
  \fmflabel{1}{i1}
   \fmflabel{2}{i2}
    \fmflabel{3}{i3}
     \fmflabel{5}{i4}
      \fmflabel{4}{i5}
\fmf{plain}{i1,v1}
\fmf{plain}{i2,v1}
\fmf{plain}{v1,v2}
  \fmf{plain,tension=0}{v2,i3}
  \fmf{dot}{v2,v3}
  \fmf{plain}{v3,i4}
   \fmf{plain}{v3,i5}
   \end{fmfgraph*}}\,\,   \quad \quad +  \parbox{20mm}{\begin{fmfgraph*}(20,20) 
  \fmftop{i1,i2}
  \fmfright{i3}
  \fmfbottom{i4,i5}
  \fmflabel{2}{i1}
   \fmflabel{3}{i2}
    \fmflabel{1}{i3}
     \fmflabel{5}{i4}
      \fmflabel{4}{i5}
\fmf{plain}{i1,v1}
\fmf{plain}{i2,v1}
\fmf{plain}{v1,v2}
  \fmf{plain,tension=0}{v2,i3}
  \fmf{dot}{v2,v3}
  \fmf{plain}{v3,i4}
   \fmf{plain}{v3,i5}
   \end{fmfgraph*}}\,\,   \quad \quad +  \parbox{20mm}{\begin{fmfgraph*}(20,20) 
  \fmftop{i1,i2}
  \fmfright{i3}
  \fmfbottom{i4,i5}
  \fmflabel{3}{i1}
   \fmflabel{1}{i2}
    \fmflabel{2}{i3}
     \fmflabel{5}{i4}
      \fmflabel{4}{i5}
\fmf{plain}{i1,v1}
\fmf{plain}{i2,v1}
\fmf{plain}{v1,v2}
  \fmf{plain,tension=0}{v2,i3}
  \fmf{dot}{v2,v3}
  \fmf{plain}{v3,i4}
   \fmf{plain}{v3,i5}
   \end{fmfgraph*}}   \,\, \quad + \quad {\rm e.\,\,\, t.}
    \end{align}
    \bigskip

\noindent We would then have for the sum of the kinematic numerators
\begin{align}\label{sum-1}
n_{12}^3 +n_{23}^1+n_{31}^2  =\qquad  \parbox{20mm}{\begin{fmfgraph*}(20,20) 
  \fmftop{i1,i2}
  \fmfright{i3}
  \fmfbottom{i4,i5}
  \fmflabel{1}{i1}
   \fmflabel{2}{i2}
    \fmflabel{3}{i3}
     \fmflabel{5}{i4}
      \fmflabel{4}{i5}
\fmf{plain}{i1,v1}
\fmf{plain}{i2,v1}
\fmf{plain}{v1,v2}
  \fmf{plain,tension=0}{v2,i3}
  \fmf{plain}{v2,v3}
  \fmf{plain}{v3,i4}
   \fmf{plain}{v3,i5}
   \end{fmfgraph*}}\,\,   \quad \quad +  \parbox{20mm}{\begin{fmfgraph*}(20,20) 
  \fmftop{i1,i2}
  \fmfright{i3}
  \fmfbottom{i4,i5}
  \fmflabel{2}{i1}
   \fmflabel{3}{i2}
    \fmflabel{1}{i3}
     \fmflabel{5}{i4}
      \fmflabel{4}{i5}
\fmf{plain}{i1,v1}
\fmf{plain}{i2,v1}
\fmf{plain}{v1,v2}
  \fmf{plain,tension=0}{v2,i3}
  \fmf{plain}{v2,v3}
  \fmf{plain}{v3,i4}
   \fmf{plain}{v3,i5}
   \end{fmfgraph*}}\,\,   \quad \quad +  \parbox{20mm}{\begin{fmfgraph*}(20,20) 
  \fmftop{i1,i2}
  \fmfright{i3}
  \fmfbottom{i4,i5}
  \fmflabel{3}{i1}
   \fmflabel{1}{i2}
    \fmflabel{2}{i3}
     \fmflabel{5}{i4}
      \fmflabel{4}{i5}
\fmf{plain}{i1,v1}
\fmf{plain}{i2,v1}
\fmf{plain}{v1,v2}
  \fmf{plain,tension=0}{v2,i3}
  \fmf{plain}{v2,v3}
  \fmf{plain}{v3,i4}
   \fmf{plain}{v3,i5}
   \end{fmfgraph*}} \\ \nonumber \\ \nonumber
+   \qquad  \parbox{20mm}{\begin{fmfgraph*}(20,20) 
  \fmftop{i1,i2}
  \fmfright{i3}
  \fmfbottom{i4,i5}
  \fmflabel{1}{i1}
   \fmflabel{2}{i2}
    \fmflabel{3}{i3}
     \fmflabel{5}{i4}
      \fmflabel{4}{i5}
\fmf{plain}{i1,v1}
\fmf{plain}{i2,v1}
\fmf{dot}{v1,v2}
  \fmf{plain,tension=0}{v2,i3}
  \fmf{plain}{v2,v3}
  \fmf{plain}{v3,i4}
   \fmf{plain}{v3,i5}
   \end{fmfgraph*}}\,\,   \quad \quad +  \parbox{20mm}{\begin{fmfgraph*}(20,20) 
  \fmftop{i1,i2}
  \fmfright{i3}
  \fmfbottom{i4,i5}
  \fmflabel{2}{i1}
   \fmflabel{3}{i2}
    \fmflabel{1}{i3}
     \fmflabel{5}{i4}
      \fmflabel{4}{i5}
\fmf{plain}{i1,v1}
\fmf{plain}{i2,v1}
\fmf{dot}{v1,v2}
  \fmf{plain,tension=0}{v2,i3}
  \fmf{plain}{v2,v3}
  \fmf{plain}{v3,i4}
   \fmf{plain}{v3,i5}
   \end{fmfgraph*}}\,\,   \quad \quad +  \parbox{20mm}{\begin{fmfgraph*}(20,20) 
  \fmftop{i1,i2}
  \fmfright{i3}
  \fmfbottom{i4,i5}
  \fmflabel{3}{i1}
   \fmflabel{1}{i2}
    \fmflabel{2}{i3}
     \fmflabel{5}{i4}
      \fmflabel{4}{i5}
\fmf{plain}{i1,v1}
\fmf{plain}{i2,v1}
\fmf{dot}{v1,v2}
  \fmf{plain,tension=0}{v2,i3}
  \fmf{plain}{v2,v3}
  \fmf{plain}{v3,i4}
   \fmf{plain}{v3,i5}
   \end{fmfgraph*}} \\ \nonumber  \\ \nonumber
 + \qquad  \parbox{20mm}{\begin{fmfgraph*}(20,20) 
  \fmftop{i1,i2}
  \fmfright{i3}
  \fmfbottom{i4,i5}
  \fmflabel{1}{i1}
   \fmflabel{2}{i2}
    \fmflabel{3}{i3}
     \fmflabel{5}{i4}
      \fmflabel{4}{i5}
\fmf{plain}{i1,v1}
\fmf{plain}{i2,v1}
\fmf{plain}{v1,v2}
  \fmf{plain,tension=0}{v2,i3}
  \fmf{dot}{v2,v3}
  \fmf{plain}{v3,i4}
   \fmf{plain}{v3,i5}
   \end{fmfgraph*}}\,\,   \quad \quad +  \parbox{20mm}{\begin{fmfgraph*}(20,20) 
  \fmftop{i1,i2}
  \fmfright{i3}
  \fmfbottom{i4,i5}
  \fmflabel{2}{i1}
   \fmflabel{3}{i2}
    \fmflabel{1}{i3}
     \fmflabel{5}{i4}
      \fmflabel{4}{i5}
\fmf{plain}{i1,v1}
\fmf{plain}{i2,v1}
\fmf{plain}{v1,v2}
  \fmf{plain,tension=0}{v2,i3}
  \fmf{dot}{v2,v3}
  \fmf{plain}{v3,i4}
   \fmf{plain}{v3,i5}
   \end{fmfgraph*}}\,\,   \quad \quad +  \parbox{20mm}{\begin{fmfgraph*}(20,20) 
  \fmftop{i1,i2}
  \fmfright{i3}
  \fmfbottom{i4,i5}
  \fmflabel{3}{i1}
   \fmflabel{1}{i2}
    \fmflabel{2}{i3}
     \fmflabel{5}{i4}
      \fmflabel{4}{i5}
\fmf{plain}{i1,v1}
\fmf{plain}{i2,v1}
\fmf{plain}{v1,v2}
  \fmf{plain,tension=0}{v2,i3}
  \fmf{dot}{v2,v3}
  \fmf{plain}{v3,i4}
   \fmf{plain}{v3,i5}
   \end{fmfgraph*}}   \,\, \quad + \quad {\rm e.\,\,\, t.}
    \end{align}
    \bigskip
  
\noindent The "exact terms" here correspond to terms where a purely longitudinal vector field appears on an undotted line of a diagram. When physical transverse vector fields are inserted into the external lines all such terms vanish, so they can be ignored for purposes of understanding of the colour-kinematics duality. 

Now comes the crucial point in our (potential) interpretation of the colour-kinematics. It is clear that what appears on the right-hand-side of (\ref{sum-1}) fails short of giving zero because terms where both intermediate lines are projected onto the ${\mathfrak u}^\perp$ are absent. Indeed, we can write

\bigskip
\be\label{sum-2}
0= n_{12}^3 +n_{23}^1+n_{31}^2  \, + \qquad  \parbox{20mm}{\begin{fmfgraph*}(20,20) 
  \fmftop{i1,i2}
  \fmfright{i3}
  \fmfbottom{i4,i5}
  \fmflabel{1}{i1}
   \fmflabel{2}{i2}
    \fmflabel{3}{i3}
     \fmflabel{5}{i4}
      \fmflabel{4}{i5}
\fmf{plain}{i1,v1}
\fmf{plain}{i2,v1}
\fmf{dot}{v1,v2}
  \fmf{plain,tension=0}{v2,i3}
  \fmf{dot}{v2,v3}
  \fmf{plain}{v3,i4}
   \fmf{plain}{v3,i5}
   \end{fmfgraph*}}\,\,   \quad \quad +  \parbox{20mm}{\begin{fmfgraph*}(20,20) 
  \fmftop{i1,i2}
  \fmfright{i3}
  \fmfbottom{i4,i5}
  \fmflabel{2}{i1}
   \fmflabel{3}{i2}
    \fmflabel{1}{i3}
     \fmflabel{5}{i4}
      \fmflabel{4}{i5}
\fmf{plain}{i1,v1}
\fmf{plain}{i2,v1}
\fmf{dot}{v1,v2}
  \fmf{plain,tension=0}{v2,i3}
  \fmf{dot}{v2,v3}
  \fmf{plain}{v3,i4}
   \fmf{plain}{v3,i5}
   \end{fmfgraph*}}\,\,   \quad \quad +  \parbox{20mm}{\begin{fmfgraph*}(20,20) 
  \fmftop{i1,i2}
  \fmfright{i3}
  \fmfbottom{i4,i5}
  \fmflabel{3}{i1}
   \fmflabel{1}{i2}
    \fmflabel{2}{i3}
     \fmflabel{5}{i4}
      \fmflabel{4}{i5}
\fmf{plain}{i1,v1}
\fmf{plain}{i2,v1}
\fmf{dot}{v1,v2}
  \fmf{plain,tension=0}{v2,i3}
  \fmf{dot}{v2,v3}
  \fmf{plain}{v3,i4}
   \fmf{plain}{v3,i5}
   \end{fmfgraph*}}    \,\quad +\,\,{\rm e.\,\, t.}
   \ee
\bigskip

\noindent This is because the sum of all the terms on the right-hand-side here is (modulo exact terms) just the pairing between the Jacobiator of $\xi_1+\xi_1^*,\xi_2+\xi_2^*,\xi_3+\xi_3^*$ and the bracket of $\xi_4+\xi_4^*,\xi_5+\xi_5^*$ 
\be
\la [\xi_1+\xi_1^*,\xi_2+\xi_2^*,\xi_3+\xi_3^*] [\xi_4+\xi_4^*,\xi_5+\xi_5^*] \ra =0,
\ee
The individual terms in (\ref{sum-2}) are various projections of this quantity on ${\mathfrak u},{\mathfrak u}^\perp$ parts of the Drinfeld double on the intermediate lines. Summing over all possible projections is equivalent to removing the projections, and so (\ref{sum-2}) follows from the usual Jacobi identity in the doubled space $D={\mathfrak u}\oplus{\mathfrak u}^\perp$. 

Now, the last 3 terms in the right-hand-side of (\ref{sum-2}) that are needed for the colour-kinematics duality to hold are not generated by the usual Feynman rules, even with our assumption that a representation giving equality sign in (\ref{property}) is possible. However, it is clear how to correct for this. Indeed, one just has to add to the YM Lagrangian a new 5-valent interaction term that generates the terms in (\ref{sum-2}). With these terms added, the equality in (\ref{sum-2}) would be the statement of colour-kinematics. To put it differently, the colour-kinematic dual set of numerators ${}^{\rm ck} n$ would be given by
\begin{align}\label{n*}
{}^{\rm ck}n_{12}^3:= n_{12}^3  \, + \qquad  \parbox{20mm}{\begin{fmfgraph*}(20,20) 
  \fmftop{i1,i2}
  \fmfright{i3}
  \fmfbottom{i4,i5}
  \fmflabel{1}{i1}
   \fmflabel{2}{i2}
    \fmflabel{3}{i3}
     \fmflabel{5}{i4}
      \fmflabel{4}{i5}
\fmf{plain}{i1,v1}
\fmf{plain}{i2,v1}
\fmf{dot}{v1,v2}
  \fmf{plain,tension=0}{v2,i3}
  \fmf{dot}{v2,v3}
  \fmf{plain}{v3,i4}
   \fmf{plain}{v3,i5}
   \end{fmfgraph*}}\qquad = \la [[[\xi_1+\xi_1^*,\xi_2+\xi_2^*],\xi_3+\xi_3^*],\xi_4+\xi_4^*], \xi_5+\xi_5^* \ra
   \\ \nonumber \\ \nonumber
 =  \qquad  \parbox{20mm}{\begin{fmfgraph*}(20,20) 
  \fmftop{i1,i2}
  \fmfright{i3}
  \fmfbottom{i4,i5}
  \fmflabel{1}{i1}
   \fmflabel{2}{i2}
    \fmflabel{3}{i3}
     \fmflabel{5}{i4}
      \fmflabel{4}{i5}
\fmf{plain}{i1,v1}
\fmf{plain}{i2,v1}
\fmf{plain}{v1,v2}
  \fmf{plain,tension=0}{v2,i3}
  \fmf{plain}{v2,v3}
  \fmf{plain}{v3,i4}
   \fmf{plain}{v3,i5}
   \end{fmfgraph*}} \qquad + \qquad \parbox{20mm}{\begin{fmfgraph*}(20,20) 
  \fmftop{i1,i2}
  \fmfright{i3}
  \fmfbottom{i4,i5}
  \fmflabel{1}{i1}
   \fmflabel{2}{i2}
    \fmflabel{3}{i3}
     \fmflabel{5}{i4}
      \fmflabel{4}{i5}
\fmf{plain}{i1,v1}
\fmf{plain}{i2,v1}
\fmf{dot}{v1,v2}
  \fmf{plain,tension=0}{v2,i3}
  \fmf{plain}{v2,v3}
  \fmf{plain}{v3,i4}
   \fmf{plain}{v3,i5}
   \end{fmfgraph*}} \qquad + \qquad \parbox{20mm}{\begin{fmfgraph*}(20,20) 
  \fmftop{i1,i2}
  \fmfright{i3}
  \fmfbottom{i4,i5}
  \fmflabel{1}{i1}
   \fmflabel{2}{i2}
    \fmflabel{3}{i3}
     \fmflabel{5}{i4}
      \fmflabel{4}{i5}
\fmf{plain}{i1,v1}
\fmf{plain}{i2,v1}
\fmf{plain}{v1,v2}
  \fmf{plain,tension=0}{v2,i3}
  \fmf{dot}{v2,v3}
  \fmf{plain}{v3,i4}
   \fmf{plain}{v3,i5}
   \end{fmfgraph*}} \qquad + \qquad \parbox{20mm}{\begin{fmfgraph*}(20,20) 
  \fmftop{i1,i2}
  \fmfright{i3}
  \fmfbottom{i4,i5}
  \fmflabel{1}{i1}
   \fmflabel{2}{i2}
    \fmflabel{3}{i3}
     \fmflabel{5}{i4}
      \fmflabel{4}{i5}
\fmf{plain}{i1,v1}
\fmf{plain}{i2,v1}
\fmf{dot}{v1,v2}
  \fmf{plain,tension=0}{v2,i3}
  \fmf{dot}{v2,v3}
  \fmf{plain}{v3,i4}
   \fmf{plain}{v3,i5}
   \end{fmfgraph*}} 
   \end{align}
 \bigskip  
   
\noindent Here $n$ are the numerators produced by the Feynman rules and the term represented by a picture in the first line is what is missing for the colour-kinematics duality to hold. Thus, we are led to the colour-kinematic dual numerators that are given by simple successive application of the bracket in the Drinfeld double. 

This suggestion of how the kinematic factors satisfying the duality could be obtained is confirmed by the findings in \cite{Bern:2010yg}, where the authors observe that in order to obtain the colour-kinematic dual numerators at 5 points, one has to add to the YM Lagrangian certain non-local term, which is a total divergence not affecting the amplitudes. We propose that the term in (\ref{n*}) that is needed to give the colour-kinematic dual numerators can similarly be interpreted as coming from a new 5-valent vertex that is to be added to the Lagrangian without affecting the amplitudes. 

\subsection{The reality}

We, however, should face the reality in which, at least for the structures described in this paper, there is no equality sign in (\ref{property}). Thus, what is described in the previous subsection is only a possibility for how the colour-kinematics could work, and for how the colour-kinematic dual numerators could be obtained. Without having the equal sign in (\ref{property}) we cannot translate the terms in the first line of (\ref{third-terms}) into objects having Lie algebraic interpretation. The only thing we can say about this sum is that it is proportional to $s_{45}$. Hence, the failure of the kinematic numerators in (\ref{n-sum}) to add up to zero is a multiple of $s_{45}$. It is then possible to see how the equations for the generalised Jacobi-like identities (3.24)-(3.27) of \cite{BjerrumBohr:2010zs} are satisfied, see the Appendix for this verification.

\section{Discussion}

Let us start by summarising what has been achieved. We have extracted from the YM Feynman rules what we called the YM bracket, which is an anti-symmetric operation that sends a pair of vector fields into a vector field. This YM bracket is closely linked to the Jacobi bracket, but does not coincide with it. We then saw that the Jacobiator of the YM bracket is cancelled by the contributions from the YM quartic vertex, apart from on-shell vanishing terms. Thus, YM Feynman rules lead to a version of the Jacobi identity at 4 points. 

We then interpreted this identity as the Jacobi identity of the Drinfeld double. The Drinfeld double is a construction that associates any Lie algebra a certain Lie algebraic structure on the sum of the Lie algebra and its dual. Having a metric in our disposal, we can associate any vector field its dual one-form. We can then twist the Drinfeld double of vector fields by considering elements of the form $\xi + \xi^*$. We then saw that the YM bracket is simply the projection of the bracket on the Drinfeld double on elements of this form. The failure of the Jacobi identity to be satisfied is then a simple consequence of the fact that elements of this form do not form a Lie sub-algebra. We also saw that the Jacobi-like identity that we discovered from the YM Feynman rules is the Jacobi identity of the Drinfeld double. 

The resulting Lie-algebraic interpretation suggests how the colour-kinematics duality can work at higher points. Thus, we were led to a very simple expression (\ref{n*}) for the would-be colour-kinematic dual numerators. These should be given simply by a successive application of the Dinfeld double bracket.

The story that we presented is only a partial success because, as we saw, to be able to convert the numerators following from Feynman rules into quantities having Lie-algebraic interpretation we need an equality sign in (\ref{property}). Thus, we need to find a version of the Feynman rules where individual contributions from the 4-valent vertex are representable as coming from a product of two cubic vertices. This is a non-trivial requirement, as is clear from inspection of these contributions. Indeed, these contributions are of the form
\be\label{v4-part}
(k_1+k_2)^2 ((\xi_1 \xi_3) (\xi_2 \xi_4) - (\xi_2 \xi_3) (\xi_1 \xi_4)).
\ee
They are clearly not representable as a product of two structures coming from some cubic vertices. Indeed, each such cubic vertex should be linear in momenta. Then, in order for the result to contain the momentum squared, the cubic vertex on vector fields $\xi_1,\xi_2$ must be of the form $k (\xi_1 \xi_2)$ for some momentum $k$. But in order for this to be anti-symmetric this must be $(k_1-k_2)(\xi_1\xi_2)$. And indeed, we saw in (\ref{v4-identity}) that the sum of contributions of the form (\ref{v4-part}) can be rewritten as the sum of squares of quantities like $(k_1-k_2)(\xi_1\xi_2)$. But does not hold for individual contributions, and this is why there is no equality sign in (\ref{property}). 

Before we discuss possible ways to go around this obstacle, let us further comment on our suggested expression for the kinematic numerators (\ref{n*}). First, these are well-defined expressions that can be computed using the bracket in the Drinfeld double. Second, by construction they satisfy the Jacobi identities and in this sense are colour-kinematic dual. However, if one attempts to use them to construct the amplitudes one will not obtain those of YM theory. This is because in the construction of the amplitudes one will be dividing the numerators by the propagators, and this produces incorrect expressions for the parts that are graphically represented by diagrams with dots on intermediate lines. This is because these diagrams will contain factors of the form $(k_1-k_3)(k_2-k_4)=s_{12}-s_{23}$. This is to be divided by $s_{13}$ when forming the amplitude. The denominator then does not cancel the numerator, and this produces wrong expressions for the amplitudes. In particular, the amplitude produced according to these rules at 4 points fails to be gauge invariant, because of the wrong contribution from the 4-valent vertex diagram.  

As far as we are aware, there is no way around this obstacle in representing (\ref{v4-part}) as a square unless one introduces some extra structure. One known to us way to obtain such a representation is specific to four space-time dimensions, and requires the introduction of the $\epsilon_{\mu\nu\rho\sigma}$ tensor into the game. Indeed, let us introduce the following bracket
\be
[\xi_1,\xi_2]^*_\mu := (\eta_{\mu\rho} \eta_{\nu\sigma} - \eta_{\mu\sigma}\eta_{\nu\rho} + \epsilon_{\mu\nu\rho\sigma}) (k_1+k_2)^\nu \xi_1^\rho \xi_2^\sigma.
\ee
Here we assumed that the signature of the metric is Euclidean. Note that this is an anti-symmetric operation on vector fields and in this sense can correspond to a "bracket". A computation gives
\be
-([\xi_1,\xi_2]^* [\xi_3,\xi_4]^*)= (k_1+k_2)^2 ( (\xi_1 \xi_3) (\xi_2 \xi_4) - (\xi_2 \xi_3) (\xi_1 \xi_4) + (\epsilon\xi_1\xi_2\xi_3\xi_4)) . 
\ee
Here $(\epsilon\xi_1\xi_2\xi_3\xi_4):=\epsilon_{\mu\nu\rho\sigma} \xi_1^\mu \xi_2^\nu \xi_3^\rho \xi_4^\sigma$.
Apart from the last term that contracts all vector fields into the $\epsilon$ tensor, this coincides with (\ref{v4-part}). It remains to be seen if this can be completed to some Drinfeld double type structure with a bracket satisfying the Jacobi identity. 

The above computation suggests that it may be possible to achieve a different representation of the Feynman rules in which there would be an equality sign in (\ref{property}) by adding an appropriate multiple of $\epsilon^{\mu\nu\rho\sigma} F^a_{\mu\nu} F^a_{\rho\sigma}$ into the action. Being a total derivative this would not change the amplitudes, but adds precisely the quantity $(\epsilon\xi_1\xi_2\xi_3\xi_4)$ to the (colour stripped) quartic vertex. Thus, adding such a term has a potential to make the story outlined in the main body of the paper work. We leave exploration of this possibility, which is specific to four dimensions, to future work. 

We finish by pointing out that there are similarities between the double structure that arose in this work, and the structures one encounters in the double field theory \cite{Hull:2009mi}, \cite{Hull:2009zb}. The later also mixes the metric and the anti-symmetric tensor similar to (\ref{X}), as well as considers objects of the type $\xi+\xi^*$. The structure of the double field theory is governed by a $D+D$ dimensional metric of signature $(D,D)$, as is the case also here. The main difference lies in the fact that the Courant bracket that is used in double field theory does not satisfy the Jacobi identity, unlike the bracket in the Drinfeld double case. However, the Courant bracket satisfies Jacobi modulo exact terms, which can be sufficient for the purposes of explaining the on-shell colour-kinematics duality. At the moment of writing we do not know whether the structure encountered in double field theory with its C-bracket has anything to do with the colour-kinematics duality of YM theory, but the similarity with the structures that arose in this paper are striking.

\section*{Acknowledgments}

The authors were supported by ERC Starting Grant 277570-DIGT\@. KK is grateful to Laurent Freidel and Djordje Minic for a stimulating discussion on the topic of this paper.

\section*{Appendix}

The purpose of this Appendix is to explicitly verify the numerator identities of \cite{BjerrumBohr:2010zs} using the formalism developed in the main body of the paper. 

\subsection{The numerator identities}

We follow the notations of \cite{BjerrumBohr:2010zs}. 
At $5$ points there are $6$ color ordered amplitudes.
Together there are $15$ different channels, and we label the residues
as $n_{1}$, $n_{2}$, $\dots$, $n_{15}$ as the following. 
\begin{eqnarray}
A(12345) & = & \frac{n_{1}}{s_{12}s_{45}}+\frac{n_{2}}{s_{23}s_{15}}+\frac{n_{3}}{s_{34}s_{12}}+\frac{n_{4}}{s_{45}s_{23}}+\frac{n_{5}}{s_{15}s_{34}}\label{eq:a12345-vh}\\
A(14325) & = & \frac{n_{6}}{s_{14}s_{25}}+\frac{n_{5}}{s_{34}s_{15}}+\frac{n_{7}}{s_{23}s_{14}}+\frac{n_{8}}{s_{25}s_{34}}+\frac{n_{2}}{s_{15}s_{23}}\nonumber\\
A(13425) & = & \frac{n_{9}}{s_{13}s_{25}}-\frac{n_{5}}{s_{34}s_{15}}+\frac{n_{10}}{s_{24}s_{13}}-\frac{n_{8}}{s_{25}s_{34}}+\frac{n_{11}}{s_{15}s_{24}}\nonumber\\
A(12435) & = & \frac{n_{12}}{s_{12}s_{35}}+\frac{n_{11}}{s_{24}s_{15}}-\frac{n_{3}}{s_{34}s_{12}}+\frac{n_{13}}{s_{35}s_{24}}-\frac{n_{5}}{s_{15}s_{34}}\nonumber\\
A(14235) & = & \frac{n_{14}}{s_{14}s_{35}}-\frac{n_{11}}{s_{24}s_{15}}-\frac{n_{7}}{s_{23}s_{14}}-\frac{n_{13}}{s_{35}s_{24}}-\frac{n_{2}}{s_{15}s_{23}}\nonumber\\
A(13245) & = & \frac{n_{15}}{s_{13}s_{45}}-\frac{n_{2}}{s_{23}s_{15}}-\frac{n_{10}}{s_{24}s_{13}}-\frac{n_{4}}{s_{45}s_{23}}-\frac{n_{11}}{s_{15}s_{24}}\nonumber
\end{eqnarray}
These are equations (3.9) to (3.14) in \cite{BjerrumBohr:2010zs}. 
We pack these residues into jacobi-like sums,
\begin{eqnarray}
X_{1} & = & n_{3}-n_{5}+n_{8}\\
X_{2} & = & n_{3}-n_{1}+n_{12}\\
X_{3} & = & n_{4}-n_{1}+n_{15}\\
X_{4} & = & n_{4}-n_{2}+n_{7}\\
X_{5} & = & n_{5}-n_{2}+n_{11}\\
X_{6} & = & n_{7}-n_{6}+n_{14}\\
X_{7} & = & n_{8}-n_{6}+n_{9}\\
X_{8} & = & n_{10}-n_{9}+n_{15}\\
X_{9} & = & n_{10}-n_{11}+n_{13}
\end{eqnarray}
Then the BCJ amplitude relations (equations
(3.15) to (3.18) in \cite{BjerrumBohr:2010zs}) are translated into the four conditions
\begin{eqnarray}
\frac{X_{3}}{s_{45}}-\frac{X_{9}}{s_{24}}-\frac{X_{2}}{s_{12}}-\frac{X_{5}}{s_{51}} & = & 0\label{eq:jacobi-like-sum-1}\\
\frac{X_{6}}{s_{14}}-\frac{X_{9}}{s_{24}}-\frac{X_{7}}{s_{25}}-\frac{X_{5}}{s_{51}} & = & 0\\
\frac{X_{8}}{s_{13}}+\frac{X_{5}}{s_{51}}-\frac{X_{4}}{s_{23}}+\frac{X_{7}}{s_{25}} & = & 0\\
\frac{X_{3}}{s_{45}}-\frac{X_{8}}{s_{13}}-\frac{X_{5}}{s_{51}}-\frac{X_{1}}{s_{34}} & = & 0\label{eq:jacobi-like-sum-4}
\end{eqnarray}
It is these equations that we would like to verify explicitly using the Drinfeld double formalism we developed. 

\subsection{Translating numerators into the language of Drinfeld double}

In the main body of the paper we have split the quartic vertex contribution into 3 parts, according to the colour structure. Each of these 3 parts was multiplied with its own Mandelstam invariant, according to how the colour contracts. For example, the $s$-channel contribution is 
\begin{equation}
s_{12}\left(\eta_{13}\eta_{24}-\eta_{14}\eta_{23}\right) \sim \begin{minipage}{1.65cm}
 \includegraphics[height=2.05cm]{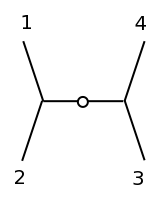}
\end{minipage}
\end{equation}
Here we also gave a convenient graphical representation. We also note that the colour-ordered quartic vertex can be expressed as the sum of two such contributions, one associated
with $s$ and one with $t$-channel respectively, 
\begin{eqnarray}
V_{\text{color-ordered}}(1234) & = & 2\eta_{13}\eta_{24}-\eta_{12}\eta_{34}-\eta_{14}\eta_{23}\\
 & = & \frac{s_{12}}{s_{12}}\left(\eta_{13}\eta_{24}-\eta_{14}\eta_{23}\right)+\frac{s_{23}}{s_{23}}\left(\eta_{13}\eta_{24}-\eta_{12}\eta_{34}\right)
\end{eqnarray}
In the main body of the paper we also introduced a different graphical notation
\begin{equation}
\begin{minipage}{1.65cm}
\includegraphics[height=2.05cm]{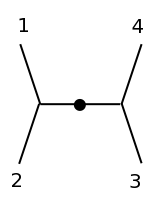}
\end{minipage} 
\sim\eta_{12}\eta_{34}(k_{1}-k_{2})\cdot(k_{3}-k_{4}),
\end{equation}
The graphs carrying black dots admit Drinfeld double interpretation, see (\ref{black-dot-interpr}). As we also explained in the main text, graphs carrying black dots are {\it not the same} as the white dotted graphs coming directly from Feynman rules, but their jacobi sums do turn out to be identical, see (\ref{identity}). We state this identity here again for convenience
\begin{eqnarray}
& & 
\begin{minipage}{5.2cm}
\includegraphics[height=2.05cm]{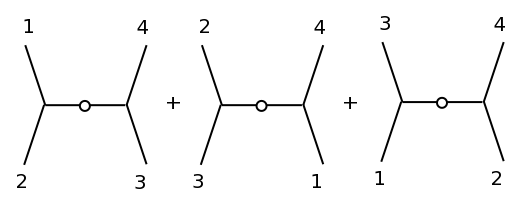}
\end{minipage}
 \nonumber \\
& & =
\begin{minipage}{5.2cm}
\includegraphics[height=2.05cm]{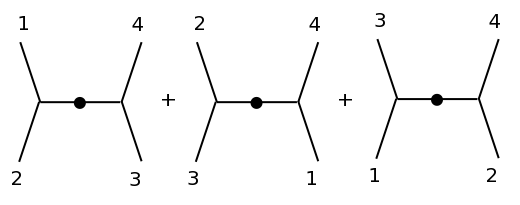}
\end{minipage}
\end{eqnarray}

\subsection{5-point amplitude}

Using the above language the $5$-point color-ordered amplitude $A(12345)$
is represented by
\begin{equation}
A(12345)= 
\begin{minipage}{8.9cm}
\includegraphics[height=6.05cm]{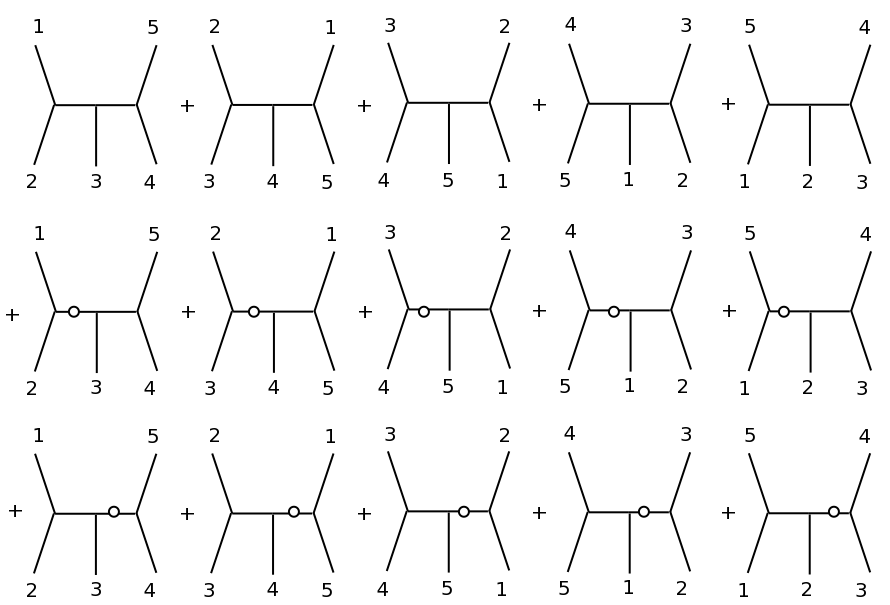}
\end{minipage}
 \label{eq:a12345}
\end{equation}
Rest of the amplitudes in equation (\ref{eq:a12345-vh}) follow similarly.
We see that the Feynman graphs are tripled (comparing to those of
a cubic theory). In addition to the original cubic graph, there is
now one graph with a white dot on the left, and one with a white dot
on the right. Comparison with equation (\ref{eq:a12345-vh}) suggests we define
\begin{equation}
n_{1}= 
\begin{minipage}{5.2cm}
\includegraphics[height=2.05cm]{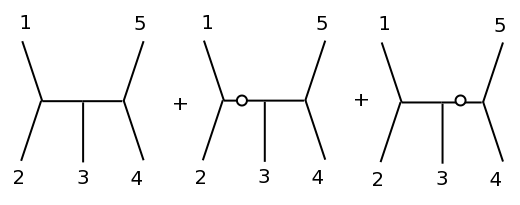}
\end{minipage}
.
\end{equation}
We know from \cite{BjerrumBohr:2010zs} that equations (\ref{eq:jacobi-like-sum-1})-(\ref{eq:jacobi-like-sum-4}) must hold since they are equivalent to BCJ amplitude relations. Our aim is to verify this explicitly. 

\subsection{Verification}

Let us first compute $X_{3}$. We have
\begin{eqnarray}
\frac{X_{3}}{s_{45}} =  \frac{n_{4}-n_{1}+n_{15}}{s_{45}}
  =  \frac{-1}{s_{45}}\,\Biggl( 
\begin{minipage}{5.2cm}
 \includegraphics[height=2.05cm]{n1}
 \end{minipage}  +123\,\text{cyclic permutations}\Biggr) \nonumber 
\end{eqnarray}
Note that graphs with white dots on the left are now in cyclic permutation
sum, which allows us to trade them with those carrying black dots
(up to longitudinal terms which vanish because they are dotted either
with polarizations or with sub-amplitudes). We therefore have
\begin{eqnarray}
\frac{X_{3}}{s_{45}} & = & \frac{-1}{s_{45}}\Biggl( 
\begin{minipage}{5.2cm}
\includegraphics[height=2.05cm]{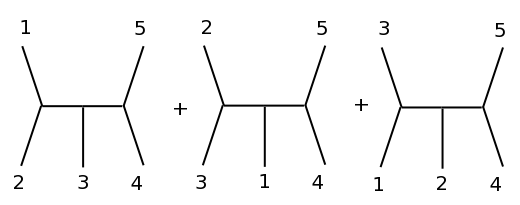}
\end{minipage} \\
 &  & + 
 \begin{minipage}{5.2cm}
 \includegraphics[height=2.05cm]{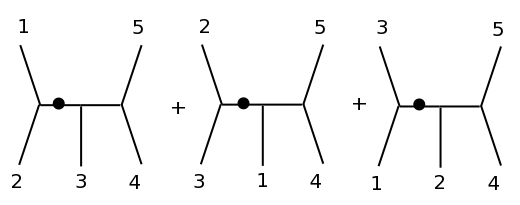}
 \end{minipage}
 \nonumber \\
 &  & +
 \begin{minipage}{5.2cm} 
 \includegraphics[height=2.05cm]{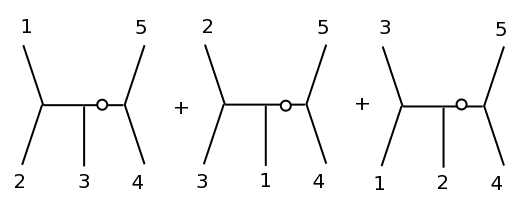}
 \end{minipage} \Biggr) \nonumber 
\end{eqnarray}
The first two lines of the above equation add up to zero because of
the jacobi identity of diffeomorphism algebra, leaving only a cyclic
sum of graphs carrying white dots on the ``wrong'' propagator

\begin{eqnarray}
\frac{X_{3}}{s_{45}} & = & \frac{-1}{s_{45}} 
\begin{minipage}{5.2cm}\includegraphics[height=2.05cm]{eqn3} \end{minipage} \\
 & = & \left[\eta_{41}\eta_{5e}-\eta_{4e}\eta_{51}\right]\times\left[\eta_{23}(k_{2}-k_{3})_{e}+\eta_{3e}(k_{3}-k_{145})_{2}+\eta_{e2}(k_{145}-k_{2})_{3}\right]+\text{cyclic perm.}\\
 & = & \eta_{41}\eta_{23}(k_{2}-k_{3})_{5}+\eta_{41}\eta_{35}(k_{3}-k_{145})_{2}+\eta_{41}\eta_{52}(k_{145}-k_{2})_{3}\label{eq:x3}\\
 &  & -\eta_{51}\eta_{23}(k_{2}-k_{3})_{4}-\eta_{51}\eta_{34}(k_{3}-k_{145})_{2}-\eta_{51}\eta_{42}(k_{145}-k_{2})_{3}+\text{cyclic perm.}\nonumber 
\end{eqnarray}
The overall $1/s_{45}$ cancels the corresponding Mandelstam variable
provided by the white dot, leaving terms linear in momentum. The same
story applies to $X_{9}$, $X_{2}$ and $X_{5}$ as well,
they are all cyclic sums of white dot graphs on the wrong propagator, so that
we can obtain their contributions directly by substituting labels,
verifying 
\begin{equation}
\frac{X_{3}}{s_{45}}-\frac{X_{9}}{s_{24}}-\frac{X_{2}}{s_{12}}-\frac{X_{5}}{s_{51}}=0.\label{eq:x3925}
\end{equation}
Explicitly this is more easily done if we concentrate on specific terms. For
example terms proportional to $\eta_{41}\eta_{23}(\text{some }k\,)_{5}$
cancel up to longitudinal terms.

   \end{fmffile}

\end{document}